\documentclass[sigconf, authorversion, nonacm, screen]{acmart}
\AtBeginDocument{%
  }

\usepackage{booktabs}
\usepackage{multicol}
\usepackage{multirow}
\usepackage[capitalize]{cleveref}
\usepackage{graphicx}
\usepackage{caption}
\usepackage{amssymb}

\usepackage{amsthm} 
\usepackage{pifont}
\usepackage{bbding}
\usepackage{amsmath,amsfonts}
\usepackage{balance}
\usepackage{fontawesome5}
\usepackage[skins,breakable]{tcolorbox}

\usepackage{tikz}
\usepackage{circuitikz}

\usepackage[shortlabels]{enumitem}
\setenumerate[1]{itemsep=0pt,partopsep=0pt,parsep=\parskip,topsep=0.5pt}
\setitemize[1]{itemsep=0pt,partopsep=0pt,parsep=\parskip,topsep=0.5pt}
\setdescription{itemsep=0pt,partopsep=0pt,parsep=\parskip,topsep=0.5pt}

\newtheorem{definition}{\bf Definition}
\newtheorem{example}{\bf Example}

\newcommand{\remarkbox}[1]{
\vspace{-\topsep}
\begin{tcolorbox}[colback=gray!10, colframe=gray!50, coltitle=black, width=\columnwidth, boxrule=0.5pt, arc=1mm, boxsep=0mm, left=1.5mm, right=1.5mm, top=1.5mm, bottom=1.5mm, breakable]
{#1}
\end{tcolorbox}
\vspace{-\topsep}
}

\newcommand{\dataset}{\textsc{nlcTables}}
\newcommand{\task}{\textsc{nlcTD}}

\newcommand{\GTR}{\textsc{Gtr}}
\newcommand{\Strubert}{\textsc{StruBert}}
\newcommand{\Santos}{\textsc{Santos}}
\newcommand{\Starmie}{\textsc{Starmie}}
\newcommand{\Josie}{\textsc{Josie}}
\newcommand{\DeepJoin}{\textsc{DeepJoin}}

\newcommand{\TUS}{\textsc{Tus}}

\begin{document}
\fancyhead{}

\newif\iffirstcall
\firstcalltrue %

\newcommand{\recommand}[2]{%
  \iffirstcall
    #1%
    \global\firstcallfalse %
  \else
    #2%
  \fi
}

\newcommand{\plainfootnote}[1]{%
  \begingroup
  \renewcommand{\thefootnote}{}%
  \footnotetext{#1}%
  \endgroup
}

\title{\recommand{\faComments~\faTable~\dataset{}: A Dataset for Marrying Natural Language Conditions with Table Discovery}{\dataset{}: A Dataset for Marrying Natural Language Conditions with Table Discovery}}

\author{
Lingxi Cui$^{1,2}$ \hspace{0.5cm} Huan Li$^{1,2}$~\Envelope \hspace{0.5cm} Ke Chen$^{1,2}$ \hspace{0.5cm} Lidan Shou$^{1,2}$~\Envelope \hspace{0.5cm} Gang Chen$^1$
\vspace{1.6mm}\\
\fontsize{10}{10}\selectfont $^1$The State Key Laboratory of Blockchain and Data Security, Zhejiang University, China\\ %
\fontsize{10}{10}\selectfont $^2$Hangzhou High-Tech Zone (Binjiang) Institute of Blockchain and Data Security, China\\ %
\fontsize{9}{9}\selectfont\ttfamily\upshape \{cuilingxi.cs, lihuan.cs, chenk, should, cg\}@zju.edu.cn
}

\renewcommand{\authors}{Lingxi Cui, Huan Li, Ke Chen, Lidan Shou, Gang Chen}
\renewcommand{\shortauthors}{Cui et al.}

\begin{abstract}

    With the growing abundance of repositories containing tabular data, discovering relevant tables for in-depth analysis remains a challenging task. 
    Existing table discovery methods primarily retrieve desired tables based on a query table or several vague keywords, leaving users to manually filter large result sets. 
    To address this limitation, we propose a new task: NL-conditional table discovery (\task{}), where users combine a query table with natural language (NL) requirements to refine search results.
    To advance research in this area, we present \dataset{}, a comprehensive benchmark dataset comprising 627 diverse queries spanning NL-only, union, join, and fuzzy conditions, 22,080 candidate tables, and 21,200 relevance annotations. 
    Our evaluation of six state-of-the-art table discovery methods on \dataset{} reveals substantial performance gaps, highlighting the need for advanced techniques to tackle this challenging \task{} scenario. 
    The dataset, construction framework, and baseline implementations are publicly available at \url{https://github.com/SuDIS-ZJU/nlcTables} to foster future research.
\end{abstract}



\maketitle

\section{Introduction}
\label{sec:intro}

\plainfootnote{The definitive Version of Record was published in SIGIR'25, \url{https://doi.org/10.1145/3726302.3730296}}

\begin{figure}[]
  \includegraphics[width=\columnwidth]{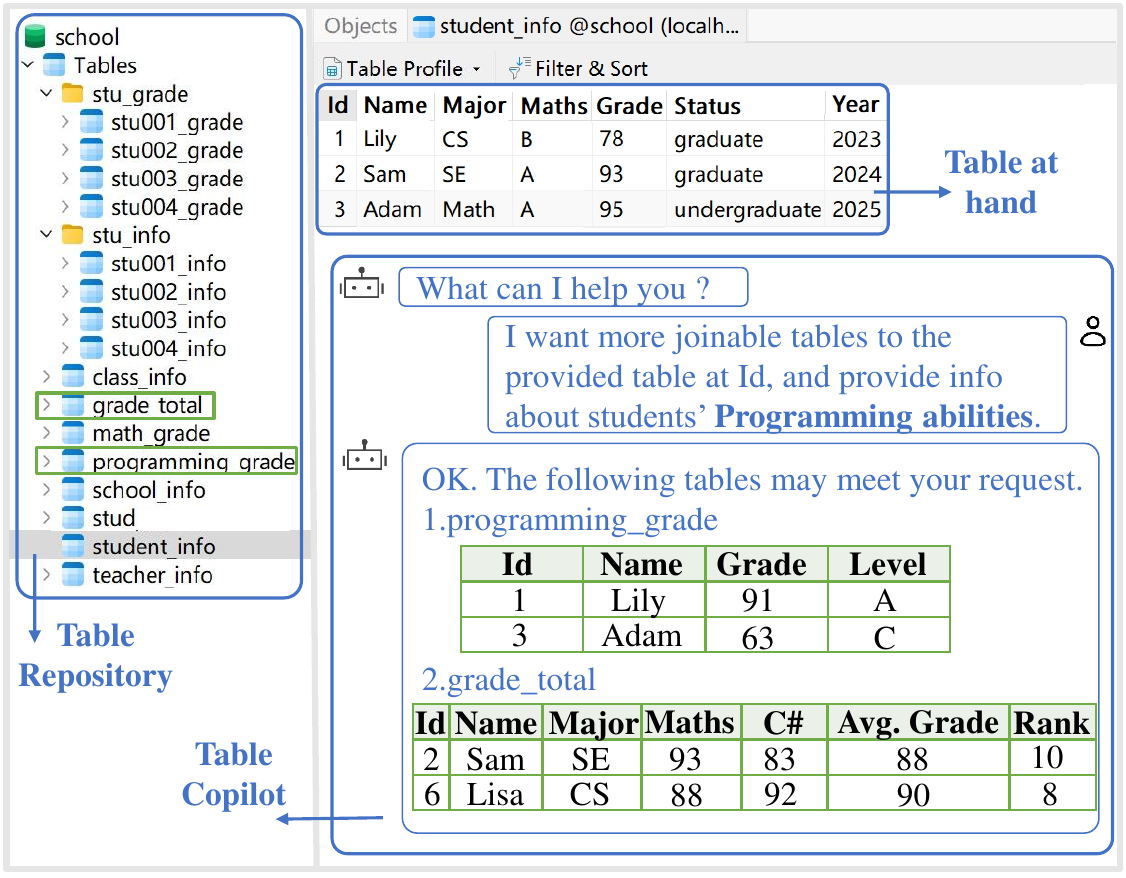}
  \caption{\textsc{Table Copilot}: a typical \task{} scenario.}
  \Description{}
  \label{fig-copilot}
\end{figure}

Tables are a widely used format for data storage, manipulation, and management. In recent decades, the availability of open and shared tables from governments, academia, and private companies has grown significantly~\cite{fan_table_2023, Web2020Shraga}, leading to an enormous number of table repositories for future analysis~\cite{paton_dataset_2024}. Table discovery~\cite{deng2024lakebench,Table2020Zhiyu,Leventidis2024ALS}, the process of identifying relevant tables, has become crucial for data users and machine learning tasks.
Figure~\ref{fig-copilot} depicts \textsc{Table Copilot}, a future paradigm of table discovery we envision, that serves as an interactive agent understanding both tables and natural language.

Currently, table discovery tasks can be generally divided into two categories: 
{keyword-based} and {query-table-based}~\cite{fan_table_2023,paton_dataset_2024,lakebench2024chai,Zhang2021SemanticTR}. 

\textbf{Keyword-based search} takes keywords as input and returns a ranked list of tables by relevance~\cite{wang_retrieving_2021,trabelsi_strubert_2022,wang_solo_2023}, allowing users to retrieve tables without prior knowledge of the repository.
However, such queries often contain only a few keywords, resulting in an overwhelming number of tables for user selection.
Meanwhile, repositories are often plagued by incomplete, meaningless, or inconsistent metadata due to the diverse origins of tables, making keyword-based search less effective~\cite{deng2024lakebench}.

\begin{figure}[!htbp]
  \includegraphics[width=\columnwidth]{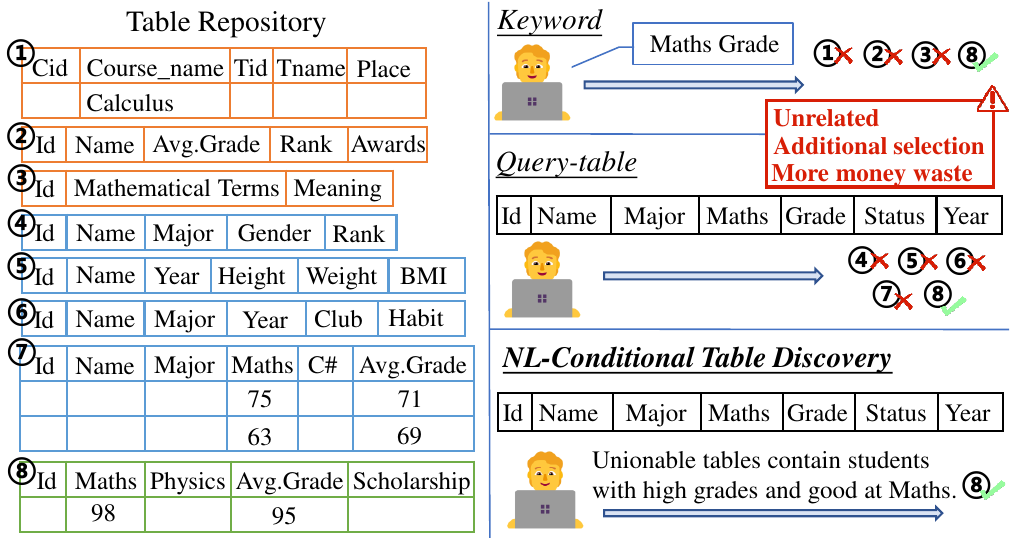}
  \caption{Illustration of NL-conditional table discovery: Combining the query table with NL conditions (e.g., high-Maths-grade students) enables more precise table retrieval.}
  \Description{}
  \label{fig-nlcTD-ex}
\end{figure}

\textbf{Query-table-based search} accepts an existing table as a query, and uses its header and/or body to locate relevant tables~\cite{paton_dataset_2024,santos_sketch-based_2022,fan_semantics-aware_2023,hu_automatic_2023}, allowing users to enhance or expand their target table (often obtained via other methods such as keyword-based search).
This field primarily focuses on two tasks: \textbf{\emph{table union search}} that finds tables with new rows to extend the query table~\cite{fan_table_2023,khatiwada_santos_2023,nargesian_table_2018,bogatu_dataset_2020,fan_semantics-aware_2023} and \textbf{\emph{table join search}} that identifies tables with new attributes to enrich the query table~\cite{dong_deepjoin_2023,zhu_josie_2019,dong_efficient_2021,chepurko_arda_2020}.
However, existing methods typically focus on either joinable or unionable tables, which suits data management experts but not general users who may lack knowledge of these distinctions yet still need additional attributes or records. 
Furthermore, simply retrieving unionable or joinable tables is often not specific enough. 
For instance, tables \ding{175}--\ding{178} in Figure~\ref{fig-nlcTD-ex} may be joinable or unionable with the query table but fail to meet the user's intent (e.g., targeting high-grade students). 
This forces users to manually filter through numerous candidates.
Similarly, keyword-based queries may return undesired results, such as tables \ding{172}–\ding{174} in Figure~\ref{fig-nlcTD-ex}, that do not relate to the query table.

To close this gap, we define a new user scenario: NL-conditional table discovery (\task{}), where users can provide a query table along with a natural language requirement. 
This setting is of great utility in real applications,
as users often have specific objectives when searching for tables (e.g., expanding a category like ``good students'' in the original table, as shown in Figure~\ref{fig-nlcTD-ex}) and specific conditions (e.g., finding students with $\mathit{grade} > 90$). 
Especially, in table repositories containing massive tables from different sources, it is an essential trend to incorporate table search assistants similar to Microsoft \textsc{Copilot}~\cite{Copilot_github,Chat2Data}.
As depicted in Figure~\ref{fig-copilot}, such tools enable users to express needs interactively, making table search more precise and user-friendly.
In this light, combining tables with conditional statements as input is a critical scenario to support.

\begin{example}\label{example:app} %
    Imagine a teacher analyzing students' performance with an existing table containing information like student ID, name, and major (as shown in Figure~\ref{fig-nlcTD-ex}).
    If he directly uses this table to search for related tables, the system might return many matches with varying content (e.g., table \ding{177} includes students' habits), making it challenging to find desired tables. 
    At this point, if a condition can be added upon the original table (such as ``I want a table that can be unioned with the original table and includes students with a high grade.''), the retrieved tables will better satisfy user needs and thus reduce user selection efforts.
\end{example}

As described in Example~\ref{example:app}, \task{} enables users to specify requirements in natural language alongside a table at hand.
This paradigm eliminates the need for extensive manual filtering or additional model training for data selection.
Its interactive nature empowers users to iteratively refine their search and adjust requirements timely.
Meanwhile, in data marketplaces where information equates to monetary value, minimizing the number of tables viewed also translates to reduced financial costs.

Recognizing the significance of this new practical scenario, we propose a novel dataset that integrates both query tables and natural language (NL) requirements. 
Unlike existing table discovery datasets, which are limited to keyword-based or query-table-based queries, our dataset \dataset{} is the first to enable conditional table search by combining query tables with NL inputs.
Additionally, we categorize the NL requirements into fine-grained subcategories to address key factors influencing the quality of conditional table searches. Specifically, our contributions are as follows:
\begin{enumerate}[leftmargin=*]
\item We propose a practical scenario, \emph{NL-conditional table discovery}, which bridges the gap between existing table discovery methods and real-world complex use cases.
We also provide a comprehensive taxonomy for this scenario, categorizing it into \textbf{16 subcategories} to systematically define its scope (Section~\ref{sec:background}).

\item We develop and release an automated and highly configurable \textbf{dataset construction framework}, including tools for generating queries, synthesizing NL requests, and annotating relevance to ensure high-quality, reproducible datasets.
Using this framework, we construct the benchmark dataset \dataset{}, which comprises: \textbf{627 realistic queries} spanning NL-only, union, join, and fuzzy conditions, \textbf{22,080 tables} from a large-scale repository covering table sizes from 1 to 69.5K rows and up to 33 columns, and \textbf{21,200 relevance annotations} with abundant positive and negative ground truths per query type (Section~\ref{sec:dataset-construct}). 

\item We evaluate \textbf{6 representative table discovery methods} on \dataset{} across various tasks, condition types, dataset compositions, and scales, confirming their limitations in addressing \task{} and revealing promising research directions (Section~\ref{sec:exp}).
\end{enumerate}

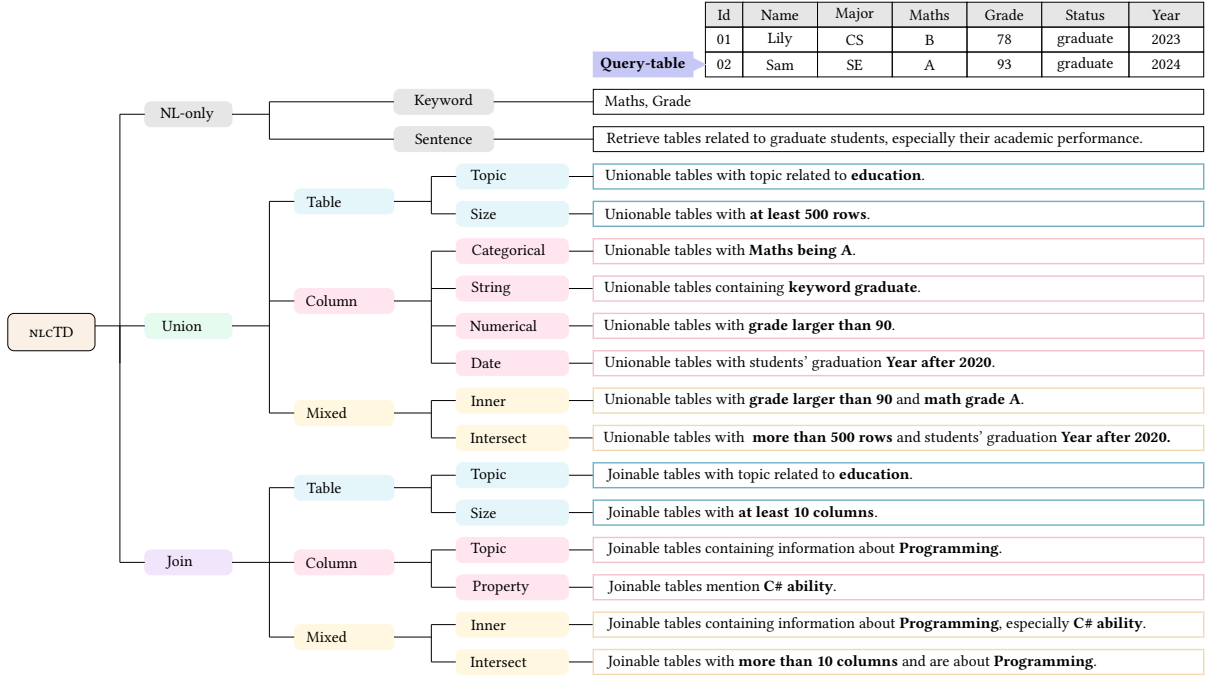
\begin{figure*}[!ht]
\centering
\resizebox{0.9\textwidth}{!}{
\begin{circuitikz}
\tikzstyle{every node}=[font=\normalsize]
\draw [ color={rgb,255:red,230; green,230; blue,230} , fill={rgb,255:red,230; green,230; blue,230}, line width=0.2pt , rounded corners = 3.6] (6,17.25) rectangle (8,16.75);
\draw [ color={rgb,255:red,230; green,230; blue,230} , fill={rgb,255:red,230; green,230; blue,230}, line width=0.2pt , rounded corners = 3.6] (6,18) rectangle (8,17.5);
\draw [ color={rgb,255:red,230; green,230; blue,230} , fill={rgb,255:red,230; green,230; blue,230}, rounded corners = 3.6] (1,17.75) rectangle (2.75,17.25);
\draw [ color={rgb,255:red,110; green,176; blue,196} , line width=0.8pt ] (10,16.5) rectangle (22.25,16);
\draw [short] (10,16.25) .. controls (9.75,16.25) and (9.75,16.25) .. (9.5,16.25);
\draw [ color={rgb,255:red,229; green,246; blue,251} , fill={rgb,255:red,229; green,246; blue,251}, rounded corners = 3.6] (7.25,16.5) rectangle (9.5,16);
\node (tikzmaker) [shift={(0.25, -0)}] at (8.25,16.75) {};
\node (tikzmaker) [shift={(1, -0)}] at (7.75,16.25) {};
\node (tikzmaker) [shift={(1, -0)}] at (7.75,15) {};
\node (tikzmaker) [shift={(1, -0)}] at (7.25,16.75) {};
\node (tikzmaker) [shift={(1, -0)}] at (7.25,16.75) {};
\node (tikzmaker) [shift={(0.25, -0.25)}] at (7.25,16.75) {};
\node (tikzmaker) [shift={(1, -0)}] at (7.25,15) {};
\draw [short] (10,15.5) .. controls (9.75,15.5) and (9.75,15.5) .. (9.5,15.5);
\node (tikzmaker) [shift={(0.25, -0)}] at (8.75,16.25) {};
\node (tikzmaker) [shift={(1, -0)}] at (8.25,16) {};
\node (tikzmaker) [shift={(1, -0)}] at (7.75,16.25) {};
\node (tikzmaker) [shift={(1, -0)}] at (7.75,16.25) {};
\node (tikzmaker) [shift={(0.25, -0.25)}] at (7.75,16.5) {};
\draw [ color={rgb,255:red,244; green,194; blue,203} , line width=0.8pt ] (10,15) rectangle (22.25,14.5);
\draw [short] (10,14.75) .. controls (9.75,14.75) and (9.75,14.75) .. (9.5,14.75);
\node (tikzmaker) [shift={(0.25, -0)}] at (8.25,15.25) {};
\node (tikzmaker) [shift={(1, -0)}] at (7.75,14.75) {};
\node (tikzmaker) [shift={(1, -0)}] at (7.75,13.5) {};
\node (tikzmaker) [shift={(1, -0)}] at (7.25,15.25) {};
\node (tikzmaker) [shift={(1, -0)}] at (7.25,15.25) {};
\node (tikzmaker) [shift={(0.25, -0.25)}] at (7.25,15.25) {};
\node (tikzmaker) [shift={(1, -0)}] at (7.25,13.5) {};
\draw [short] (10,14) .. controls (9.75,14) and (9.75,14) .. (9.5,14);
\node (tikzmaker) [shift={(0.25, -0)}] at (8.75,14.75) {};
\node (tikzmaker) [shift={(1, -0)}] at (8.25,14.5) {};
\node (tikzmaker) [shift={(1, -0)}] at (7.75,14.75) {};
\node (tikzmaker) [shift={(1, -0)}] at (7.75,14.75) {};
\node (tikzmaker) [shift={(0.25, -0.25)}] at (7.75,15) {};
\draw [short] (10,13.25) .. controls (9.75,13.25) and (9.75,13.25) .. (9.5,13.25);
\node (tikzmaker) [shift={(0.25, -0)}] at (8.25,13.75) {};
\node (tikzmaker) [shift={(1, -0)}] at (7.75,13.25) {};
\node (tikzmaker) [shift={(1, -0)}] at (7.75,12) {};
\node (tikzmaker) [shift={(1, -0)}] at (7.25,13.75) {};
\node (tikzmaker) [shift={(1, -0)}] at (7.25,13.75) {};
\node (tikzmaker) [shift={(0.25, -0.25)}] at (7.25,13.75) {};
\node (tikzmaker) [shift={(1, -0)}] at (7.25,12) {};
\draw [short] (10,12.5) .. controls (9.75,12.5) and (9.75,12.5) .. (9.5,12.5);
\node (tikzmaker) [shift={(0.25, -0)}] at (8.75,13.25) {};
\node (tikzmaker) [shift={(1, -0)}] at (8.25,13) {};
\node (tikzmaker) [shift={(1, -0)}] at (7.75,13.25) {};
\node (tikzmaker) [shift={(1, -0)}] at (7.75,13.25) {};
\node (tikzmaker) [shift={(0.25, -0.25)}] at (7.75,13.5) {};
\draw [ color={rgb,255:red,242; green,218; blue,176} , line width=0.8pt ] (10,12) rectangle (22.25,11.5);
\draw [short] (10,11.75) .. controls (9.75,11.75) and (9.75,11.75) .. (9.5,11.75);
\node (tikzmaker) [shift={(0.25, -0)}] at (8.25,12.25) {};
\node (tikzmaker) [shift={(1, -0)}] at (7.75,11.75) {};
\node (tikzmaker) [shift={(1, -0)}] at (7.75,10.5) {};
\node (tikzmaker) [shift={(1, -0)}] at (7.25,12.25) {};
\node (tikzmaker) [shift={(1, -0)}] at (7.25,12.25) {};
\node (tikzmaker) [shift={(0.25, -0.25)}] at (7.25,12.25) {};
\node (tikzmaker) [shift={(1, -0)}] at (7.25,10.5) {};
\draw [short] (10,11) .. controls (9.75,11) and (9.75,11) .. (9.5,11);
\node (tikzmaker) [shift={(0.25, -0)}] at (8.75,11.75) {};
\node (tikzmaker) [shift={(1, -0)}] at (8.25,11.5) {};
\node (tikzmaker) [shift={(1, -0)}] at (7.75,11.75) {};
\node (tikzmaker) [shift={(1, -0)}] at (7.75,11.75) {};
\node (tikzmaker) [shift={(0.25, -0.25)}] at (7.75,12) {};
\draw [short] (10,10.25) .. controls (9.75,10.25) and (9.75,10.25) .. (9.5,10.25);
\node (tikzmaker) [shift={(0.25, -0)}] at (8.25,10.75) {};
\node (tikzmaker) [shift={(1, -0)}] at (7.75,10.25) {};
\node (tikzmaker) [shift={(1, -0)}] at (7.75,9) {};
\node (tikzmaker) [shift={(1, -0)}] at (7.25,10.75) {};
\node (tikzmaker) [shift={(1, -0)}] at (7.25,10.75) {};
\node (tikzmaker) [shift={(0.25, -0.25)}] at (7.25,10.75) {};
\node (tikzmaker) [shift={(1, -0)}] at (7.25,9) {};
\draw [short] (10,9.5) .. controls (9.75,9.5) and (9.75,9.5) .. (9.5,9.5);
\node (tikzmaker) [shift={(0.25, -0)}] at (8.75,10.25) {};
\node (tikzmaker) [shift={(1, -0)}] at (8.25,10) {};
\node (tikzmaker) [shift={(1, -0)}] at (7.75,10.25) {};
\node (tikzmaker) [shift={(1, -0)}] at (7.75,10.25) {};
\node (tikzmaker) [shift={(0.25, -0.25)}] at (7.75,10.5) {};
\draw [short] (10,8.75) .. controls (9.75,8.75) and (9.75,8.75) .. (9.5,8.75);
\node (tikzmaker) [shift={(0.25, -0)}] at (8.25,9.25) {};
\node (tikzmaker) [shift={(1, -0)}] at (7.75,8.75) {};
\node (tikzmaker) [shift={(1, -0)}] at (7.75,7.5) {};
\node (tikzmaker) [shift={(1, -0)}] at (7.25,9.25) {};
\node (tikzmaker) [shift={(1, -0)}] at (7.25,9.25) {};
\node (tikzmaker) [shift={(0.25, -0.25)}] at (7.25,9.25) {};
\node (tikzmaker) [shift={(1, -0)}] at (7.25,7.5) {};
\draw [short] (10,8) .. controls (9.75,8) and (9.75,8) .. (9.5,8);
\node (tikzmaker) [shift={(0.25, -0)}] at (8.75,8.75) {};
\node (tikzmaker) [shift={(1, -0)}] at (8.25,8.5) {};
\node (tikzmaker) [shift={(1, -0)}] at (7.75,8.75) {};
\node (tikzmaker) [shift={(1, -0)}] at (7.75,8.75) {};
\node (tikzmaker) [shift={(0.25, -0.25)}] at (7.75,9) {};
\draw [short] (10,7.25) .. controls (9.75,7.25) and (9.75,7.25) .. (9.5,7.25);
\node (tikzmaker) [shift={(0.25, -0)}] at (8.25,7.75) {};
\node (tikzmaker) [shift={(1, -0)}] at (7.75,7.25) {};
\node (tikzmaker) [shift={(1, -0)}] at (7.25,7.75) {};
\node (tikzmaker) [shift={(1, -0)}] at (7.25,7.75) {};
\node (tikzmaker) [shift={(0.25, -0.25)}] at (7.25,7.75) {};
\draw [short] (10,6.5) .. controls (9.75,6.5) and (9.75,6.5) .. (9.5,6.5);
\node (tikzmaker) [shift={(0.25, -0)}] at (8.75,7.25) {};
\node (tikzmaker) [shift={(1, -0)}] at (8.25,7) {};
\node (tikzmaker) [shift={(1, -0)}] at (7.75,7.25) {};
\node (tikzmaker) [shift={(1, -0)}] at (7.75,7.25) {};
\node (tikzmaker) [shift={(0.25, -0.25)}] at (7.75,7.5) {};
\node (tikzmaker) [shift={(0.25, -0)}] at (8.25,6.25) {};
\node (tikzmaker) [shift={(1, -0)}] at (7.25,6.25) {};
\node (tikzmaker) [shift={(1, -0)}] at (7.25,6.25) {};
\node (tikzmaker) [shift={(0.25, -0.25)}] at (7.25,6.25) {};
\draw [ color={rgb,255:red,255; green,229; blue,240} , fill={rgb,255:red,255; green,229; blue,240}, rounded corners = 3.6] (4,14) rectangle (6,13.5);
\draw [ color={rgb,255:red,229; green,246; blue,251} , fill={rgb,255:red,229; green,246; blue,251}, line width=0.2pt , rounded corners = 3.6] (4,16) rectangle  (6,15.5);
\node [font=\normalsize] at (5.5,13) {};
\draw [ color={rgb,255:red,255; green,247; blue,224} , fill={rgb,255:red,255; green,247; blue,224}, rounded corners = 3.6] (4,11.75) rectangle (6,11.25);
\draw [ color={rgb,255:red,255; green,229; blue,240} , fill={rgb,255:red,255; green,229; blue,240}, rounded corners = 3.6] (4,8.75) rectangle (6,8.25);
\draw [ color={rgb,255:red,229; green,246; blue,251} , fill={rgb,255:red,229; green,246; blue,251}, rounded corners = 3.6] (4,10.25) rectangle (6,9.75);
\node [font=\normalsize] at (5.5,6.25) {};
\draw [ color={rgb,255:red,255; green,247; blue,224} , fill={rgb,255:red,255; green,247; blue,224}, rounded corners = 3.6] (4,7.25) rectangle (6,6.75);
\draw (6.75,16.25) to[short] (7.25,16.25);
\draw (6.75,15.5) to[short] (7.25,15.5);
\draw (6.75,14.75) to[short] (7.25,14.75);
\draw (6.75,14) to[short] (7.25,14);
\draw (6.75,13.25) to[short] (7.25,13.25);
\draw (6.75,12.5) to[short] (7.25,12.5);
\node [font=\normalsize] at (5.5,8.5) {};
\draw (6.75,11.75) to[short] (7.25,11.75);
\draw (6.75,11) to[short] (7.25,11);
\draw (6.75,10.25) to[short] (7.25,10.25);
\draw (6.75,9.5) to[short] (7.25,9.5);
\draw (6.75,8.75) to[short] (7.25,8.75);
\draw (6.75,8) to[short] (7.25,8);
\draw (6.75,7.25) to[short] (7.25,7.25);
\draw (6.75,6.5) to[short] (7.25,6.5);
\draw (6.75,16.25) to[short] (6.75,15.5);
\draw (6.75,14.75) to[short] (6.75,12.5);
\draw (6.75,11.75) to[short] (6.75,11);
\draw (6.75,10.25) to[short] (6.75,9.5);
\draw (6.75,8.75) to[short] (6.75,8);
\draw (6.75,7.25) to[short] (6.75,6.5);
\draw (6,15.75) to[short] (6.75,15.75);
\draw (6,13.75) to[short] (6.75,13.75);
\draw (6,11.5) to[short] (6.75,11.5);
\draw (6,10) to[short] (6.75,10);
\draw (6,8.5) to[short] (6.75,8.5);
\draw (6,7) to[short] (6.75,7);
\draw [ color={rgb,255:red,229; green,246; blue,251} , fill={rgb,255:red,229; green,246; blue,251}, rounded corners = 3.6] (7.25,15.75) rectangle (9.5,15.25);
\draw [ color={rgb,255:red,229; green,246; blue,251} , fill={rgb,255:red,229; green,246; blue,251}, rounded corners = 3.6] (7.25,10.5) rectangle (9.5,10);
\draw [ color={rgb,255:red,229; green,246; blue,251} , fill={rgb,255:red,229; green,246; blue,251}, rounded corners = 3.6] (7.25,9.75) rectangle (9.5,9.25);
\draw [ color={rgb,255:red,255; green,229; blue,240} , fill={rgb,255:red,255; green,229; blue,240}, rounded corners = 3.6] (7.25,15) rectangle (9.5,14.5);
\draw [ color={rgb,255:red,255; green,229; blue,240} , fill={rgb,255:red,255; green,229; blue,240}, rounded corners = 3.6] (7.25,14.25) rectangle (9.5,13.75);
\draw [ color={rgb,255:red,255; green,229; blue,240} , fill={rgb,255:red,255; green,229; blue,240}, rounded corners = 3.6] (7.25,13.5) rectangle (9.5,13);
\draw [ color={rgb,255:red,255; green,229; blue,240} , fill={rgb,255:red,255; green,229; blue,240}, rounded corners = 3.6] (7.25,12.75) rectangle (9.5,12.25);
\draw [ color={rgb,255:red,255; green,229; blue,240} , fill={rgb,255:red,255; green,229; blue,240}, rounded corners = 3.6] (7.25,9) rectangle (9.5,8.5);
\draw [ color={rgb,255:red,255; green,229; blue,240} , fill={rgb,255:red,255; green,229; blue,240}, rounded corners = 3.6] (7.25,8.25) rectangle (9.5,7.75);
\draw [ color={rgb,255:red,255; green,247; blue,224} , fill={rgb,255:red,255; green,247; blue,224}, rounded corners = 3.6] (7.25,12) rectangle (9.5,11.5);
\draw [ color={rgb,255:red,255; green,247; blue,224} , fill={rgb,255:red,255; green,247; blue,224}, rounded corners = 3.6] (7.25,11.25) rectangle (9.5,10.75);
\draw [ color={rgb,255:red,255; green,247; blue,224} , fill={rgb,255:red,255; green,247; blue,224}, rounded corners = 3.6] (7.25,7.5) rectangle (9.5,7);
\draw [ color={rgb,255:red,255; green,247; blue,224} , fill={rgb,255:red,255; green,247; blue,224}, rounded corners = 3.6] (7.25,6.75) rectangle (9.5,6.25);
\draw [ color={rgb,255:red,110; green,176; blue,196} , line width=0.8pt ] (10,15.75) rectangle (22.25,15.25);
\draw [ color={rgb,255:red,110; green,176; blue,196} , line width=0.8pt ] (10,10.5) rectangle (22.25,10);
\draw [ color={rgb,255:red,110; green,176; blue,196} , line width=0.8pt ] (10,9.75) rectangle (22.25,9.25);
\draw [ color={rgb,255:red,244; green,194; blue,203} , line width=0.8pt ] (10,14.25) rectangle (22.25,13.75);
\draw [ color={rgb,255:red,244; green,194; blue,203} , line width=0.8pt ] (10,13.5) rectangle (22.25,13);
\draw [ color={rgb,255:red,244; green,194; blue,203} , line width=0.8pt ] (10,12.75) rectangle (22.25,12.25);
\draw [ color={rgb,255:red,244; green,194; blue,203} , line width=0.8pt ] (10,9) rectangle (22.25,8.5);
\draw [ color={rgb,255:red,244; green,194; blue,203} , line width=0.8pt ] (10,8.25) rectangle (22.25,7.75);
\draw [ color={rgb,255:red,242; green,218; blue,176} , line width=0.8pt ] (10,11.25) rectangle (22.25,10.75);
\draw [ color={rgb,255:red,242; green,218; blue,176} , line width=0.8pt ] (10,7.5) rectangle (22.25,7);
\draw [ color={rgb,255:red,242; green,218; blue,176} , line width=0.8pt ] (10,6.75) rectangle (22.25,6.25);
\node [font=\normalsize] at (13.45,16.25) {Unionable tables with topic related to \textbf{education}.};
\node [font=\normalsize] at (12.9,15.5) {Unionable tables with \textbf{at least 500 rows}.};
\node [font=\normalsize] at (12.75,14.75) {Unionable tables with \textbf{Maths being A}.};
\node [font=\normalsize] at (13.4,14) {Unionable tables containing \textbf{keyword graduate}.};
\node [font=\normalsize] at (13.14,13.25) {Unionable tables with \textbf{grade larger than 90}.};
\node [font=\normalsize] at (14.15,12.5) {Unionable tables with students' graduation \textbf{Year after 2020}.};
\node [font=\normalsize] at (14.44,11.75) {Unionable tables with \textbf{grade larger than 90} and \textbf{math grade A}.};
\node [font=\normalsize] at (15.9,11) {Unionable tables with \textbf{   more than 500 rows} and students' graduation \textbf{Year after 2020.}};

\node [font=\normalsize] at (13.35,10.25) {Joinable tables with topic related to \textbf{education}.};
\node [font=\normalsize] at (13,9.5) {Joinable tables with \textbf{at least 10 columns}.};
\node [font=\normalsize] at (14.25,8.75) {Joinable tables containing information about \textbf{Programming}.};
\node [font=\normalsize] at (12.6,8) {Joinable tables mention \textbf{C\# ability}.};
\node [font=\normalsize] at (15.7,7.25) {Joinable tables containing information about \textbf{Programming}, especially \textbf{C\# ability}.};
\node [font=\normalsize] at (15.2,6.5) {Joinable tables with \textbf{more than 10 columns} and are about \textbf{Programming}.};

\node [font=\normalsize] at (7.9,16.25) {Topic};
\node [font=\normalsize] at (7.8,15.5) {Size};
\node [font=\normalsize] at (8.3,14.75) {Categorical};
\node [font=\normalsize] at (7.95,14) {String};
\node [font=\normalsize] at (8.2,13.25) {Numerical};
\node [font=\normalsize] at (7.85,12.5) {Date};
\node [font=\normalsize] at (7.9,11.75) {Inner};
\node [font=\normalsize] at (8.1,11) {Intersect};
\node [font=\normalsize] at (7.9,10.25) {Topic};
\node [font=\normalsize] at (7.8,9.5) {Size};
\node [font=\normalsize] at (7.9,8.75) {Topic};
\node [font=\normalsize] at (8.15,8) {Property};
\node [font=\normalsize] at (7.9,7.25) {Inner};
\node [font=\normalsize] at (8.1,6.5) {Intersect};

\node [font=\normalsize] at (4.6,15.75) {Table};
\node [font=\normalsize] at (4.75,13.75) {Column};
\node [font=\normalsize] at (4.65,11.5) {Mixed};
\node [font=\normalsize] at (4.6,10) {Table};
\node [font=\normalsize] at (4.75,8.5) {Column};
\node [font=\normalsize] at (4.65,7) {Mixed};

\node [font=\normalsize] at (1.85,17.5) {NL-only};
\node [font=\normalsize] at (7,17.75) {Keyword};
\node [font=\normalsize] at (7,17) {Sentence};
\node [font=\normalsize] at (15.65,17) {Retrieve tables related to graduate students, especially their academic performance.};

\draw [ color={rgb,255:red,229; green,246; blue,251} , fill={rgb,255:red,229; green,251; blue,237}, line width=0.2pt , rounded corners = 3.6] (1,13.5) rectangle (2.75,13);
\node [font=\normalsize] at (1.75,13.25) {Union};
\draw [ color={rgb,255:red,241; green,229; blue,251} , fill={rgb,255:red,241; green,229; blue,251}, line width=0.2pt , rounded corners = 3.6] (1,8.75) rectangle  (2.75,8.25);
\node [font=\normalsize] at (1.7,8.5) {Join};
\draw (3.5,15.75) to[short] (4,15.75);
\draw (3.5,13.75) to[short] (4,13.75);
\draw (3.5,11.5) to[short] (4,11.5);
\draw (3.5,10) to[short] (4,10);
\draw (3.5,8.5) to[short] (4,8.5);
\draw (3.5,7) to[short] (4,7);
\draw (3.5,15.75) to[short] (3.5,11.5);
\draw (3.5,10) to[short] (3.5,7);
\draw (2.75,8.5) to[short] (3.5,8.5);
\draw (2.75,13.25) to[short] (3.5,13.25);
\draw (0.5,13.25) to[short] (1,13.25);
\draw (0.5,8.5) to[short] (1,8.5);
\draw (0.5,13.25) to[short] (0.5,8.5);
\draw [ fill={rgb,255:red,251; green,240; blue,229} , line width=0.2pt , rounded corners = 3.6] (-1.75,13.5) rectangle  node {\normalsize \task{}} (0,12.75);
\node [font=\normalsize] at (0.5,10.75) {  };
\draw  (10,17.25) rectangle (22.25,16.75);
\draw  (10,18) rectangle (22.25,17.5);
\draw [short] (8,17.75) -- (10,17.75);
\draw [short] (8,17) -- (10,17);
\draw [short] (3.5,17.75) -- (6,17.75);
\draw [short] (3.5,17) -- (6,17);
\draw [short] (3.5,17.75) -- (3.5,17);
\draw [short] (2.75,17.5) -- (3.5,17.5);
\draw [short] (0.5,17.5) -- (1,17.5);
\draw [short] (0.5,17.5) -- (0.5,12.5);
\node [font=\normalsize] at (11.1,17.75) {Maths, Grade};

\node [font=\normalsize] at (14.75,19.5) {};
\draw [short] (0,13.25) -- (0.5,13.25);

\draw [ fill={rgb,255:red,230; green,230; blue,230} ] (12.25,19.75) rectangle  node {\normalsize Id} (13,19.25);
\draw [ fill={rgb,255:red,230; green,230; blue,230} ] (13,19.75) rectangle  node {\normalsize Name} (14.5,19.25);
\draw [ fill={rgb,255:red,230; green,230; blue,230} ] (14.5,19.75) rectangle  node {\normalsize Major} (16,19.25);
\draw [ fill={rgb,255:red,230; green,230; blue,230} ] (16,19.75) rectangle  node {\normalsize Maths} (17.5,19.25);
\draw [ fill={rgb,255:red,230; green,230; blue,230} ] (17.5,19.75) rectangle  node {\normalsize Grade} (19,19.25);
\draw [ fill={rgb,255:red,230; green,230; blue,230} ] (19,19.75) rectangle  node {\normalsize Status} (20.75,19.25);
\draw [ fill={rgb,255:red,230; green,230; blue,230} ] (20.75,19.75) rectangle  node {\normalsize Year} (22.25,19.25);
\draw  (12.25,19.25) rectangle  node {\normalsize 01} (13,18.75);
\draw  (13,19.25) rectangle  node {\normalsize Lily} (14.5,18.75);
\draw  (14.5,19.25) rectangle  node {\normalsize CS} (16,18.75);
\draw  (16,19.25) rectangle  node {\normalsize B} (17.5,18.75);
\draw  (17.5,19.25) rectangle  node {\normalsize 78} (19,18.75);
\draw  (19,19.25) rectangle  node {\normalsize graduate} (20.75,18.75);
\draw  (20.75,19.25) rectangle  node {\normalsize 2023} (22.25,18.75);
\draw  (12.25,18.75) rectangle  node {\normalsize 02} (13,18.25);
\draw  (13,18.75) rectangle  node {\normalsize Sam} (14.5,18.25);
\draw  (14.5,18.75) rectangle  node {\normalsize SE} (16,18.25);
\draw  (16,18.75) rectangle  node {\normalsize A} (17.5,18.25);
\draw  (17.5,18.75) rectangle  node {\normalsize 93} (19,18.25);
\draw  (20.75,18.75) rectangle  node {\normalsize 2024} (22.25,18.25);
\draw  (19,18.75) rectangle  node {\normalsize graduate} (20.75,18.25);

\draw [ color={rgb,255:red,199; green,200; blue,245} , fill={rgb,255:red,199; green,200; blue,245}] (11.25,18.5) -- (11.75,18.25) -- (12.25,18.5) -- (11.75,18.75) -- cycle;
\draw [ color={rgb,255:red,199; green,200; blue,245} , fill={rgb,255:red,199; green,200; blue,245}] (10,18.75) rectangle (12,18.25);
\node [font=\normalsize] at (11,18.5) {\textbf{Query-table}};
\end{circuitikz}
}
\caption{The taxonomy of \task{}, consisting of 16 NL condition subcategories along with their illustrative examples.}
\label{fig:taxonomy}
\end{figure*}

\section{Task Definition and Taxonomy}
\label{sec:background}

Section~\ref{ssec:definition} gives the formal definition of NL-conditional table discovery and Section~\ref{ssec:taxonomy} further categorizes it at finer granularity.

\subsection{Definition and Scenarios}
\label{ssec:definition}

The task of NL-conditional table discovery (\task{}) is to rank and retrieve a set of top-$k$ tables from a repository that are relevant to both a query table and a natural language (NL) condition. This means that the user query is given as 1) a table organized in rows and columns, representing the user's initial dataset of interest, and 2) an NL condition, the textual specification describing extra requirements or constraints.
For example, as shown in Figure~\ref{fig-nlcTD-ex}, a user may want to find tables related to a given student table at hand and specifically filter for ``high-Maths-grade students''.
The ranking process is based not only on structural or exact matches between the query table and the candidate tables but also on the semantic relevance to the NL request. 
We formally define \task{} as follows:

\begin{definition}[NL-conditional Table Discovery]
\label{definition:nlcTD}
    Given a table repository $\mathcal{T}$, and a user query $Q$ consisting of a query table $T^q$ and an NL condition $C$, the \task{} task aims to retrieve from $\mathcal{T}$ a top-$k$ ranked list of tables $\mathcal{T}' = \{ T_i \}$ that are semantically relevant to both $T^q$ and $C$, as determined by a relevance scoring function, $\rho(T^q, C, T_i)$.
\end{definition} 

The result of an \task{} query is a ranked list of tables from the repository $\mathcal{T}$ that match the user's requirements expressed through both $T^q$ and $C$.
This setting aligns closely with real-world demands and supports a wealth of applications:
From a \textbf{user perspective}, \task{} provides an intuitive, interactive interface for data discovery, allowing users to express detailed requirements in natural language.
This makes the system accessible to non-technical users, reflecting the broader trend of NL-based data interaction nowadays~\cite{Chat2Data}.
From a \textbf{technical perspective}, \task{} enhances table discovery by enabling more targeted retrieval, particularly beneficial for table augmentation and improving the quality of AI training datasets~\cite{Cui2024TabularDA}.
It also complements technologies like \emph{Retrieval-Augmented Generation} (RAG) by providing contextually relevant and domain-specific data~\cite{tablerag_2024_Chen,IR4RAG2024}.
Furthermore, the NL condition $C$ can vary widely depending on practical scenarios. It may specify aspects such as table size, attributes, and records. To reflect this diversity, we categorize these NL conditions at finer granularity, forming the basis for constructing a comprehensive dataset for \task{}.

\subsection{Taxonomy of \task{}}
\label{ssec:taxonomy}

Figure~\ref{fig:taxonomy} provides an overview of the \task{} taxonomy with illustrative examples of NL conditions.
We begin by treating keyword-based table search as a simplified case of \task{}, extending it to form a single category.
Next, we extend query-table-based search by adding NL conditions, creating two advanced categories: 
NL-conditional table union search targets rows, while NL-conditional table join search focuses on identifying relevant columns.
Each category features distinct NL requests.
Furthermore, we classify NL conditions based on table granularity into three levels: table-level, column-level, and mixed-mode conditions.

\smallskip
\noindent\textbf{NL-only Table Search}.
This category extends traditional keyword-based table search, which relies exclusively on keywords, to include both keywords and \emph{complete sentences} for table retrieval.
By incorporating this scenario, \task{} bridges previous work and enhances the practicality and versatility of table copilot applications.

\smallskip
\noindent\textbf{NL-conditional Table Union Search}.
Table union operation involves appending tuples from one table to another that share common columns.
The objective of \emph{table union search} is to identify tables unionable with a query table, thereby enriching it with additional tuples (rows). 
Since this process {aims at introducing} new tuples, the associated NL requirements primarily impose constraints on tuples. 
For example, given a student information table as the query table (see Figure~\ref{fig:taxonomy}), union requests may specify conditions, such as ``including only students' graduation year after 2020''. 
Based on table granularity, we further classify NL-conditional table union search into the following three types.
\begin{enumerate}[leftmargin=*]
\item \underline{\textit{Table-level Conditions}} focus on global properties of the candidate tables that are unionable. They can be related to key aspects such as 
(a) Table Topic: constraints on the subject matter, e.g., requiring tables related to education, and
(b) Table Size: requirements on the number of rows or columns, e.g., tables with more than 500 rows.

\item \underline{\textit{Column-level Conditions}} impose constraints on rows in specific columns of the candidate tables. These conditions can be classified based on column type:
(a) Categorical Columns: values required to belong to specific categories or sets of categories;
(b) String Columns: presence of particular {words or} phrases, e.g., requiring ``graduate'' to appear in the column;
(c) Numerical Columns: values being greater than or less than a threshold, or averages meeting specific criteria; and
(d) Date Columns: Similar to numerical columns but with additional constraints based on date formats (e.g., ``\texttt{yyyy-mm-dd}'' or ``\texttt{dd/mm/yyyy}'').

\item \underline{\textit{Mixed-mode Conditions}} combine both table-level and column-level constraints.
These may involve intra-class mixing (e.g., table-level conditions combining table topic and table size) or inter-class mixing (e.g., a table-level condition on table size combined with a column-level condition on numerical columns). 
In practice, NL requests often contain multiple conditions, making mixed-mode conditions essential for handling complex real-world scenarios.
\end{enumerate}

\smallskip
\noindent\textbf{NL-conditional Table Join Search}.
A join operation occurs when two tables share a common column, i.e., a column in table A and a column in table B have a significant overlap in values, enabling one table's columns to be appended to the other.
The objective of \emph{table join search} is to locate relevant tables that can be joined to enrich the query table with supplementary attributes. 
Since this process aims at new attributes, the corresponding NL requirements primarily focus on specifying the desired characteristics of these new attributes. 
For example, given a student information table as the query table (see Figure~\ref{fig:taxonomy}), an NL condition for table join search might specify that the added columns should relate to a particular topic, such as Programming. 
Similar to table union search, three subcategories are divided based on table granularity.

\underline{\textit{Table-level Conditions}} and \underline{\textit{Mixed-mode Conditions}} in table join search are similar to the counterparts in table union search.
The primary distinction lies in \underline{\textit{Column-level Conditions}}: Unlike union operations, which involve adding rows and may impose constraints on data distribution, join operations do not alter the records in the query table --- they simply add attributes. Consequently, join conditions focus on the \emph{topic or property} of the new columns rather than thresholds or distributions. For example, column-level conditions might require the discovered tables to include specific types of columns, such as those containing Programming scores or other domain-specific attributes (e.g., BMI for students).

\remarkbox{
\textbf{Taxonomy Completeness}.
Our taxonomy builds upon established principles from existing table discovery methods in keyword-based and query-table-based fashions while extending them to accommodate flexible NL conditions.
It is inherently extensible. The methodology described in this study for constructing the dataset can be easily adapted to catalog new NL conditions or application-specific requirements.
In summary, it is designed to cover the vast majority of application scenarios and provides a clear pathway for integrating new conditions, ensuring its completeness and utility both now and in the future.
}

\section{Related Work}
\label{sec:related}

Section~\ref{ssec:existing_table_discovery} goes through current table discovery methods, highlighting the need and opportunities for developing \task{} techniques. Section~\ref{ssec:existing_data_collection} examines existing test data collections, identifying gaps that must be addressed to facilitate \task{}.

\subsection{Table Discovery Approaches}
\label{ssec:existing_table_discovery}

\noindent\textbf{Keyword-based Table Search}~\cite{wang_retrieving_2021,trabelsi_strubert_2022,wang_solo_2023} retrieves a ranked list of table instances from the repository, ordered by their relevance scores in relation to one or more user-provided keywords. 
These approaches allow users to locate data with minimal prior knowledge of the structure or relationships within the table repository. However, they typically accept sorely keywords as input and treat tables as plain text (e.g., \textsc{Strubert}~\cite{trabelsi_strubert_2022} serializes tables to a sequence of tokens). 
While keyword-based approaches are highly valuable, there remains significant room for improvement. \task{} defined in Section~\ref{ssec:definition} aims to extend this paradigm by supporting inputs that go beyond simple keywords, allowing for more complex natural language queries, such as full sentences. Nonetheless, taking only NL as input is only a special case in our \task{} scenarios. In practice, users often already have a table at hand and would use it as query context to refine their table retrieval results.

\smallskip
\noindent\textbf{Query-table-based Table Search} identifies additional tables relevant to a given query table, leveraging its header and/or body content.
The literature has mainly seen 
two categories of work. %

\underline{\textit{Table Union Search}}~\cite{nargesian_table_2018,bogatu_dataset_2020,khatiwada_santos_2023,fan_semantics-aware_2023,hu_automatic_2023} retrieves tables that are union-compatible with the query table, meaning these two tables having multiple pairs of columns that can be merged. This involves assessing similarity between query table and candidate tables. Early approach \TUS{}~\cite{nargesian_table_2018} defines three probabilistic models to measure the similarity between column values, column domains, and word embeddings of column content. \Santos{}~\cite{khatiwada_santos_2023} utilizes knowledge graph to measure the possibility that columns originate from the same domain.
Recently, methods like \Starmie{}~\cite{fan_semantics-aware_2023} have adopted table representation learning~\cite{deng_turl_2020,hu_automatic_2023} to generate column embeddings for similarity measure, allowing a more nuanced understanding of table contexts.

\underline{\textit{Table Join Search}}~\cite{yakout_infogather_2012,zhu_josie_2019,dong_efficient_2021,dong_deepjoin_2023,liu_feature_2022,chepurko_arda_2020,Table2022dong} retrieves tables that are joinable with the query table at a specified column $C_q$, meaning the two tables sharing overlapping or semantically related values in that column.
\Josie{}~\cite{zhu_josie_2019} considers only exact value overlap using set similarity search.
Here, $C_q$ is treated as a set, and the top-$k$ columns with the highest value overlap are returned.
Recent embedding-based approaches such as \DeepJoin{}~\cite{dong_deepjoin_2023} consider semantic overlap (e.g., ``\texttt{Incorporation}'' and ``\texttt{Inc.}'') through representation learning. Typically, they first encode the columns, then index these columns, and finally search joinable pairs based on the index.

Query-table-based search is commonly used to identify table components that can augment or complement an existing query table. 
While \task{} encompasses this functionality --- allowing users to specify conditions like retrieving unionable or joinable tables ---\task{} goes beyond these by supporting more flexible and customizable NL conditions.
Not limited to predefined operations (e.g., set similarity or semantic encoding), \task{} allows users to specify nuanced requests in natural language. Existing approaches struggle with \task{} as they fail to capture and utilize the rich semantics of these conditions (see our empirical study in Section~\ref{ssec:overall_comparison}).

\subsection{Existing Test Data Collections}
\label{ssec:existing_data_collection}

To the best of our knowledge, no prior work has specifically constructed a dataset for NL-conditional table discovery, despite its significant application potential.
Existing tabular data collections primarily focus on table retrieval (or table discovery) in either keyword-based or query-table-based paradigms.
Table~\ref{related_work_statistic} summarizes representative datasets, highlighting their focus, dataset scales, and the availability of ground truth annotations for benchmarking.
For example, \textsc{WikiTables}~\cite{zhang_ad_2018,wang_retrieving_2021,trabelsi_strubert_2022} is centered on keyword-based search, accepting only keywords in queries. 
\TUS{}~\cite{nargesian_table_2018} and \Santos{}~\cite{khatiwada_santos_2023} focus on table union search scenarios, where query tables are provided as input and ground truth is generated via table splitting (see Section~\ref{ssec:construction}).
\textsc{LakeBench}\cite{deng2024lakebench} generates table union and join search datasets by applying table splitting to \textsc{Opendata}\cite{opendata} and \textsc{WebTable}~\cite{venetis_recovering_2011,cafarella2009data}. It provides large-scale datasets aligned with data lake usage, emphasizing algorithm efficiency and scalability.

Despite these advancements, all of the aforementioned table discovery datasets only label keyword-table or table-table relatedness, making them unsuitable for tasks requiring a ``triangular'' relationship among the NL condition, the query table, and the candidate table.
This ternary relationship is inherently more complex than binary relationships, and thus necessitates the creation of a substantial and diverse experimental dataset.
While the last four datasets from \textsc{LakeBench}~\cite{deng2024lakebench} are designed to progressively increase in scale, %
the other datasets remain relatively small and are derived from single data sources. This lack of diversity makes them less suitable for scenarios involving highly varied NL conditions.
Our dataset, \dataset{}, addresses these gaps by aligning with prior works' input data levels while introducing a comprehensive table discovery framework driven by NL conditions. By supporting diverse NL inputs alongside query tables, \dataset{} enables robust and scalable evaluations for real-world applications.

\begin{table}[]
\centering
\caption{Characteristics of existing test collections. Types \texttt{K}, \texttt{U}, and \texttt{J} denote keyword-based, table union search, and table join search, respectively. The scale of each dataset is represented by the number of queries $|\{T^q\}|$, number of candidate tables $|\mathcal{T}|$, total number of ground truth \#(GT), and the row counts per table Avg \#(Rows). Numbers in \textit{italics} indicate statistics not directly reported in the original papers but derived from our evaluations of the dataset.}
\label{related_work_statistic}
\footnotesize

\begin{tabular}{lrrrrr}
\toprule
{Dataset}        & {Type} & {$|\{T^q\}|$} & {$|\mathcal{T}|$} & {\#(GT)} & {Avg \#(Rows)} \\ \midrule
\textsc{WikiTables}     & \texttt{K}    & 60      & 3K     & 3K      & 10.9     \\ 
\TUS{}      & \texttt{U}    & \textit{92}     & 5K     & \textit{5K}     & 1.9K     \\ 
\Santos{}  & \texttt{U}    & 80      & 11K    & \textit{1.6K}   & 7.7K     \\ 
\textsc{OpenData}-$\textsf{U}$       & \texttt{U} & 4.6K    & 65K   & \textit{49.5K}    & 112.4K    \\ 
\textsc{OpenData}-$\textsf{J}$ & \texttt{J} & 4.8K    & 65K    & \textit{42.6K}      & 112.4K   \\ 
\textsc{WebTable}-$\textsf{U}$       & \texttt{U} & 6.8K   & 2.8M   & \textit{54.9K}  & 23.5       \\ 
\textsc{WebTable}-$\textsf{J}$ & \texttt{J} & 7.5K   & 16.6M  & \textit{54.8K}   & 23.5     \\ \bottomrule
\end{tabular}
\end{table}

\begin{figure*}[ht]
  \includegraphics[width=\textwidth]{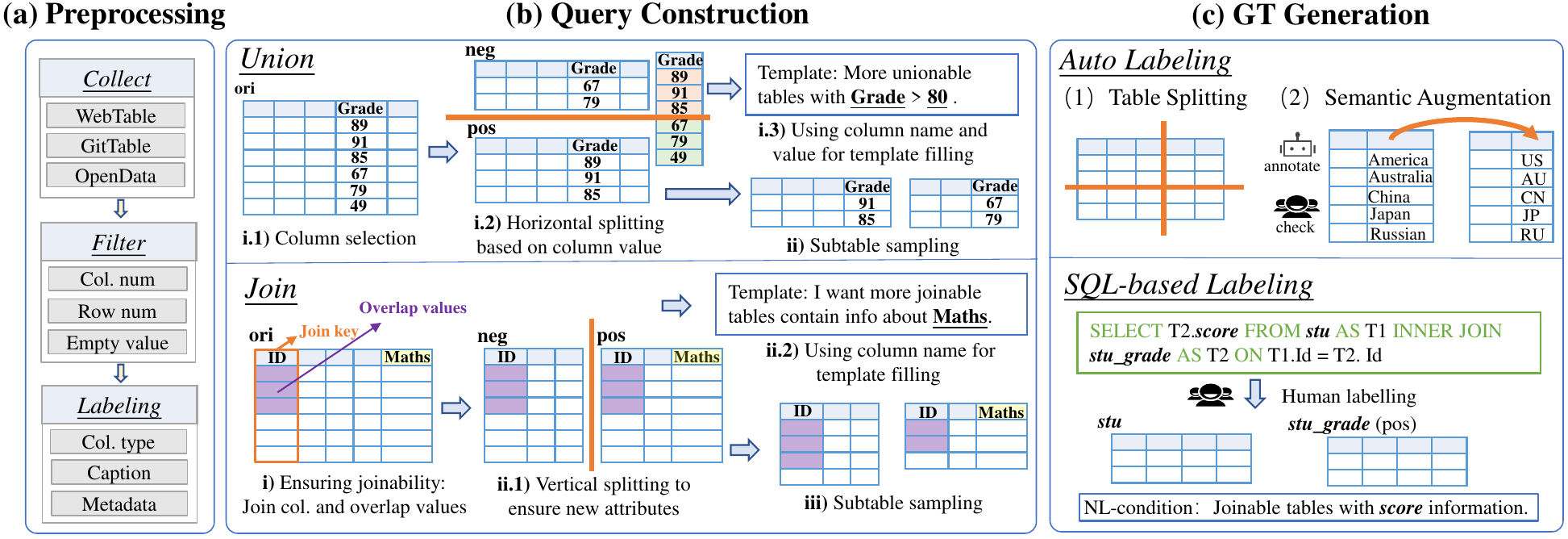}
  \caption{The three stages of constructing \dataset{}: (1) Table Preprocessing: collecting, filtering, and labeling tables; (2) Query Construction: splitting tables vertically and horizontally to create joinable and unionable tables; (3) Ground Truth Generation: generating labels via automatic table splitting with semantic augmentation, and manual SQL-based labeling.}
  \Description{}
  \label{dataset-construction}
\end{figure*}

\section{The \dataset{} Dataset}
\label{sec:dataset-construct}

Section~\ref{ssec:principles} outlines the design principles of \dataset{}, while Section~\ref{ssec:construction} describes its construction framework. Section~\ref{statistics} provides the statistics of the datasets used for evaluations.

\subsection{Design Principles}
\label{ssec:principles}

\dataset{} is designed as a comprehensive dataset for NL-conditional table discovery, guided by three core principles:

\smallskip
\noindent\textbf{P1 (Scalability)}.
A scalable dataset reduces overfitting, improves model generalization, and ensures reliability across diverse applications. To achieve this, each query should include multiple positive and negative ground truths, and large-scale table repositories (candidate tables) must be incorporated. These repositories should be assessed based on both the number of tables and their average sizes, emphasizing real-world scenarios with thousands of diverse tables or more, including those larger ones.

\smallskip
\noindent\textbf{P2 (Diversity)}. 
Query performance varies drastically depending on query types and candidate table properties. 
For instance, when comparing two queries: one targeting a {string} column and the other a numerical column, semantic-aware algorithms, like those based on pre-trained language models, are typically more effective for {string} columns.
To enable thorough evaluation, the dataset must include diverse queries and tables sourced from different platforms. The dataset should also represent queries across different categories, as defined in the taxonomy in Section~\ref{ssec:taxonomy}.

\smallskip
\noindent\textbf{P3 (Configurability)}. 
To accommodate varying user needs and query or candidate table characteristics, the dataset should be highly configurable. Key hyperparameters, such as the number of positive and negative ground truths per query, must be adjustable to allow users to tailor the dataset for specific requirements.

\subsection{Dataset Construction Framework}
\label{ssec:construction}

As depicted in Figure~\ref{dataset-construction}, the construction process consists of three main stages.
First, we collect a large and diverse set of tables and apply filtering to obtain high-quality original tables. 
Next, we adopt \emph{table splitting} to construct queries that include both NL conditions and query tables, while simultaneously generating ground truth labels.
Finally, to enhance the diversity and authenticity of the dataset, we apply large language models (LLMs) for semantic augmentation of the ground truths that have been generated via table splitting. 
Meanwhile, we manually annotate several ground truths based on real SQL use cases contained in the \textsc{Spider}~\cite{li2024can} dataset.

\smallskip
\noindent\textbf{(1) Table Preprocessing}.
We begin by collecting tables from multiple publicly accessible sources to ensure compliance with P2 (Diversity).
As shown in Figure~\ref{dataset-construction}(a), these sources include \textsc{Opendata}~\cite{opendata} (relatively larger tables provided), \textsc{WebTable}~\cite{venetis_recovering_2011,cafarella2009data} (smaller but diverse tables provided), and \textsc{GitTable}~\cite{hulsebos2023gittables} (domain-specific table with programming details provided).
To ensure the quality and utility of the data (see P1 (Scalability)), we apply strict filtering criteria. 
A table is retained only if it exceeds predefined thresholds for rows ($\geq 50$) and columns ($\geq 8$). 
This filtering aligns with the requirements of table splitting, as larger tables are more suitable for generating meaningful queries. Additionally, we remove tables with excessive empty cells or meaningless values (e.g., ``not recognized'').
After preprocessing, we retain a total of 10,755 tables, of which 193 high-quality tables are selected as the original tables for query construction.
To prepare for NL condition generation, we manually annotate the column names and types, ensuring P2 by enabling different splitting methods for various column types.

\smallskip
\noindent\textbf{(2) Query Construction}.
We adopt a table splitting approach inspired by previous datasets~\cite{nargesian_table_2018,khatiwada_santos_2023,deng2024lakebench,zhang_entitables_2017}.
The basic idea is to partition large tables into multiple small ones, with these resulting small tables considered \emph{related} (e.g, unionable and joinable). 
These smaller tables serve as query tables and candidate tables, enabling P1 (Scalability) by generating numerous ground truth pairs. Below, we describe the processes for constructing union and join queries.

\underline{\textit{Union Queries}}.  
Union queries focus on retrieving tuples (rows) that satisfy specific conditions (e.g., ``$\mathit{Grade} > 80 $''). The process includes two key operations:
(i) \textit{Horizontal Splitting} based on Specific Conditions: 
Horizontal splitting creates positive and negative ground truth tables by applying row-level conditions. For example, in Figure~\ref{dataset-construction}(b), the positive table (\texttt{pos\_GT}) satisfies ``$\mathit{Grade} > 80 $'', while the negative table does not.
For \emph{table-level conditions}, subtables from the same table naturally share the same topic (often related to the table caption which we use in our NL condition template), and we use a hyperparameter ($\texttt{L\_scale}$) to control table size thresholds.
For \emph{column-level conditions}, we select a column ($\$\texttt{col}$) from the table (e.g., categorical, string, numerical, or date) and automatically determine a specific value ($\texttt{c\_value}$) based on the column's distribution (e.g., a class for categorical columns, a number for numerical columns). The NL condition is then formed using $\$\texttt{col}$ and $\texttt{c\_value}$ (e.g., ''$\$\texttt{col}$ > $\texttt{c\_value}$'' for numerical columns). The table is split into subsets that meet or fail to meet the condition.
(ii) \textit{Subtable Sampling}: 
To ensure high-quality ground truth, we sample rows such that $80\%$ of rows in the positive table satisfy the condition, while $50\%$ of rows in the negative table do not. Multiple ground truth tables are generated by repeating this process based on the $\texttt{pos\_num}$ and $\texttt{neg\_num}$ hyperparameters, supporting P1 (Scalability).
Unlike traditional database unions that require strict schema consistency, our approach relaxes it for challenging test cases.
A pair of tables is considered unionable if they share several columns from the same domain. To simulate such realistic scenarios, we randomly select shared columns and sample additional columns as supplementary features, supporting P2 (Diversity).

\underline{\textit{Join Queries}}.
Join queries focus on retrieving columns (attributes) related to a query table's topic or properties (e.g., ``related to Programming''). The process involves three key operations:
(i) \textit{Ensuring Joinability}:
A pair of tables is considered joinable if they share at least one column with substantial overlap in cell values. 
To achieve this, we randomly select a column as the join key and ensure overlap by sampling shared rows between the query table and the ground truth table. Additional rows are added to preserve variability.
(ii) \textit{Vertical Table Splitting}:
Except for the join key, the remaining table is vertically partitioned into two subsets: one for the query table and the other for the ground truth.
This ensures that new attributes are added to the ground truth table. 
For \emph{table-level conditions}, similar to union queries, topic is naturally consistent due to table splitting; and a hyperparameter $\texttt{L\_col}$ is used to control scale thresholds. 
For \emph{column-level conditions}, a random, new column $\$\texttt{new\_col}$ is selected and referenced in the NL template (e.g., ``contains information about $\$\texttt{new\_col}$'').
(iii) \textit{Subtable Sampling}:
Similar to union queries, subtable sampling ensures multiple ground truths are generated, contributing to P1 (Scalability).

\underline{\textit{Mixed-mode Queries}}.  
We first generate one query table with corresponding positive and negative ground truths, then apply a second query only to the positive ground truths to create multiple final positive ground truths satisfying both queries. 
The negative ground truths generated from the two steps are all collected as the final negative ground truths. 
This two-step process supports both intersect mix between table-level and column-level, and inner mix within themselves\footnote{In theory, this method applies to mixing more than two conditions. However, given the current task's complexity, \dataset{} opts to support the mixing of two conditions.}, enhancing P2 (Diversity) via random combination.

\underline{\textit{NL-only Queries}}. 
We repurpose unionable and joinable queries by manually removing references to the query table in the NL condition while maintaining fluency. Adjustments are made to the ground truths to align with the modified NL conditions.

\begin{table*}[ht]
\caption{
Summary statistics of \dataset{}, including dataset types (\textsf{K}, \textsf{U}, \textsf{J}, and \textsf{fz} represent keyword-only, table union, table join, and fuzzy versions, respectively), queries, table repository, and labeled ground truths (GT). The table reports the numbers of queries, candidate tables, and GT per dataset type, with detailed ranges and mean value ($\mu$) of NL condition length, table rows, and table columns. The last column shows the ratio between positive and negative samples in GTs.}
\label{table:statistic}
\resizebox{\textwidth}{!}{
\begin{tabular}{lccccccccccl}
\toprule
\multirow{2}{*}{\textbf{Dataset Type}} & \multicolumn{5}{c}{\textbf{Queries}}                                                   & \multicolumn{3}{c}{\textbf{Table Repository}}     & \multicolumn{2}{c}{\textbf{GT}}            \\ \cmidrule(lr){2-6} \cmidrule(lr){7-9} \cmidrule(lr){10-11}
                                       & \textbf{Total} & \textbf{Tab:Col:Mix} & \textbf{Condition Length} & \textbf{Query Table Rows} & \textbf{Query Table Columns} & \textbf{Total} & \textbf{Table Rows}       & \textbf{Table Columns}     & \textbf{Total} & \textbf{Pos:Neg} \\ \midrule
\dataset{}\textsf{\_K}                 & 235            & $\smallsetminus$                    & 2$\sim$23 ($\mu$=13.9)         & $\smallsetminus$                         & $\smallsetminus$                           & 7,405          & 1$\sim$69.5K ($\mu$=285.0)     & 1$\sim$32 ($\mu$=7.4)           & 6,841          & 1:8.6            \\
\dataset{}\textsf{-U}                  & 255            & 1:7:7                & 10$\sim$68 ($\mu$=19.5)        & 4$\sim$13.4K ($\mu$=1,613.0)   & 2$\sim$25 ($\mu$=8.8)            & 7,567          & 1$\sim$69.5K ($\mu$=292.1)     & 1$\sim$32 ($\mu$=7.4)           & 7,411          & 1:10.7           \\
\dataset{}\textsf{-U-fz}               & 39             & 1:4:5                & 10$\sim$30 ($\mu$=19.3)        & 6$\sim$100 ($\mu$=82.2)        & 3$\sim$19 ($\mu$=7.2)            & 1,620          & 1$\sim$7,169 ($\mu$=2,043.5)   & 1$\sim$20 ($\mu$=5.9)           & 1,560          & 1:11.7           \\
\dataset{}\textsf{-J}                  & 91             & 3:4:8                & 9$\sim$27 ($\mu$=16.4)         & 20$\sim$5.5K ($\mu$=1,723.0)   & 3$\sim$24 ($\mu$=6.8)            & 4,871          & 1$\sim$68.5K ($\mu$=838.1)     & 2$\sim$33 ($\mu$=6.7)           & 4,821          & 1:21.8           \\
\dataset{}\textsf{-J-fz}               & 27             & 1:1:2                & 11$\sim$26 ($\mu$=16.9)        & 20$\sim$2.0K ($\mu$=134.2)     & 3$\sim$13 ($\mu$=6.3)            & 617            & 1$\sim$9.9K ($\mu$=480.9)      & 2$\sim$20 ($\mu$=6.7)           & 567            & 1:7.5            \\ \midrule
\dataset{}\textsf{-Full}      & 647            & $\smallsetminus$                    & 2$\sim$68 ($\mu$=16.9)                        & 4$\sim$5.5K ($\mu$=1395.5)                         &   2$\sim$25 ($\mu$=8.0)                         & 22,080         & 1$\sim$69.5K ($\mu$=543.9)                         & 1$\sim$33 ($\mu$=7.1)                          & 21,200         & $\smallsetminus$                \\ \bottomrule
\end{tabular}%
}
\end{table*}

\smallskip
\noindent\textbf{(3) Labeling Queries}.
As shown in Figure~\ref{dataset-construction}(c), ground truths are derived from the following two sources.

\underline{\textit{Automated Labeling}}.
During the table-splitting process, tables generated from the same large table as the query table are automatically labeled as ground truths. To enhance universality and align with P2 (Diversity), we leverage LLMs for semantic augmentation.
This involves transforming entire columns into their semantically similar counterparts (e.g., ``\texttt{Apple}'' to ``\texttt{Apple Inc.}'' and ``\texttt{China}'' to ``\texttt{CN}'') within the segmented small tables. The augmented tables are then manually reviewed to ensure quality and consistency.
Also, LLMs are used to rewrite NL conditions for linguistic diversity. 

\underline{\textit{SQL-based Manual Labeling}}.
We also derive ground truths from real-world use cases, specifically leveraging queries from the \textsc{Spider} dataset~\cite{li2024can}.
Using join and union SQL statements in conjunction with table metadata, we identify pairs of related tables: a query table and its ground truth. 
NL conditions are then carefully crafted to reflect the query context, ensuring precision and relevance.

\remarkbox{
\textbf{Principle Achievement.}
Our construction ensures that P1 (Scalability) is achieved via large-scale table splitting and sampling, P2 (Diversity) is addressed via diverse data sources, query and condition types, and LLM-based transformation, and P3 (Configurability) is supported through adjustable hyperparameters and condition composition.
}

\subsection{Dataset Statistics}
\label{statistics}

Table~\ref{table:statistic} presents an overview of \dataset{}, which includes queries (a query table paired with an NL condition), candidate tables, and labeled ground truths (GTs).
GTs explicitly indicate correct and incorrect candidate tables for each query, ensuring robust evaluation capabilities.
\dataset{} is highly configurable, with a set of adjustable hyperparameters (e.g., $\texttt{dup\_rate}$ for overlap proportion during splitting, $\texttt{temp\_num}$ for the number of NL templates). The dataset scale can be modified by varying the size of original tables or the values of $\texttt{pos\_num}$ and $\texttt{neg\_num}$.

\dataset{} supports NL-only table search (\dataset{}\textsf{\_K}), NL-conditional table union search (\dataset{}\textsf{-U}), and NL-conditional table join search (\dataset{}\textsf{-J}). 
For union and join tasks, \textbf{fuzzy versions} (\dataset{}\textsf{-U-fz} and \dataset{}\textsf{-J-fz}) are provided using semantic augmentation.
In total, \dataset{} contains 22,080 tables with large average size and includes 21,200 labeled GTs.
The dataset scale is adjustable (we provide \textsf{Small}, \textsf{Medium}, and \textsf{Large} versions of \dataset{} in Section~\ref{ssec:benchmark_composition} for dataset scale experiments), though its current size already poses significant challenges for existing table discovery methods.
Queries span across various categories, encompassing table-level, column-level, and mixed-mode conditions. Although the dataset's refined classification proportions are not fully detailed in the paper due to page limit, they can adapt dynamically to the attributes of the original tables (e.g., whether a table is large-scale or includes numerical or date columns).

The construction of \dataset{} demonstrates meticulous design and effort: Over 12,947 lines of code were written to automate preprocessing, table splitting, query generation, and semantic augmentation.
The team dedicated approximately 168 hours to manual annotation, covering table preprocessing, SQL-based GT labeling, semantic augmentation review, and NL-only query writing.

\section{Evaluation on \dataset{}}
\label{sec:exp}

We evaluate various table discovery methods on the newly constructed \dataset{} to address the following research questions.

\smallskip
\textbf{RQ1}. \emph{How do prior table discovery methods perform in the new \task{} scenarios in terms of effectiveness?} (Section~\ref{ssec:overall_comparison})

\noindent\textbf{Short Answer}: Existing keyword-based and query-table-based methods perform poorly in this new scenario, as they are not designed to handle the query table and the NL condition together.

\smallskip
\textbf{RQ2}. \emph{How do existing table discovery methods perform on different categories of NL conditions?} (Section~\ref{ssec:condition_categories})

\noindent\textbf{Short Answer}: 
 Queries on table topics perform better, likely due to alignment with keywords, while numerical and date columns are frequently neglected as current methods lack sensitivity to column attributes. 
 Mixed-mode conditions are highly challenging due to their complexity. 
These highlight significant limitations of existing methods, especially in handling nuanced and diverse NL conditions.

\smallskip
\textbf{RQ3}. \emph{How robust are these methods, especially when modifying the composition of the dataset?} (Section~\ref{ssec:benchmark_composition})

\noindent\textbf{Short Answer}: Most methods show reduced accuracy as the number of negative ground truths increases, indicating a lack of robustness in distinguishing relevant tables from irrelevant ones. When we increase the size of datasets, those without a special index structure suffer. Furthermore, when we augment the dataset with \emph{fuzzy} queries, most methods experience a performance drop, particularly those that are not semantically sensitive. 

\subsection{Experimental Settings}
\label{ssec:setup}

We implement all experiments in Python and execute them on a DELL server equipped with a single Intel Xeon w9-3495X CPU (56 cores, 4.8 GHz) and two NVIDIA RTX A6000 GPUs. All experiments are conducted in the same runtime to ensure fair comparisons.

\smallskip  
\noindent\textbf{Metric}. 
Following prior works~\cite{nargesian_table_2018,khatiwada_santos_2023,fan_semantics-aware_2023,wang_retrieving_2021,dong_deepjoin_2023,zhang_table2vec_2019}, we evaluate table discovery methods using Precision$@k$ ($P@k$), Recall$@k$ ($R@k$) and Normalized Discounted Cumulative Gain$@k$ ($\mathit{NDCG}@k$).
Formally, for a query $Q = (T^q, C)$ (see Definition~\ref{definition:nlcTD}) and the set $\mathcal{T}'$ of top-$k$ tables retrieved by a method, $\mathcal{T}^g$ the ground truth tables, and $\rho_i$ the relevance score of the $i$-th retrieved table in $\mathcal{T}'$.
The metrics are defined as:
$P@k = {|\mathcal{T}^g \cap \mathcal{T}'|}/{|\mathcal{T}'|}$,
$R@k = {|\mathcal{T}^g \cap \mathcal{T}'|}/{|\mathcal{T}^g|}$, and
$\mathit{NDCG}@k = {1}/{Z_k} \sum_{i=1}^{k} {\rho_i}/{\log_2(i+1)}$,
where $Z_k$ is a 0-1 normalization factor.
For each method, we report the average $P@k$, $R@k$, and $\mathit{NDCG}@k$ across all queries.

\smallskip
\noindent\textbf{Baselines}. 
We include six representative approaches published at top-tier venues, two each for keyword-based search (No.~1--2), table union search (No.~3--4), and table join search (No.~5--6):
\begin{enumerate}[leftmargin=*]

\item \GTR{}~\cite{wang_retrieving_2021} (\emph{SIGIR}'21), a keyword-based table retrieval method that converts tables into tabular graphs (cell, row, and column nodes) and uses a Graph Transformer to capture both content and structural layout. Finally, it performs query-table matching over \textsc{Bert}-embedded keyword-like queries.

\item \Strubert{}~\cite{trabelsi_strubert_2022} (\emph{WWW}'22),
a keyword-based table retrieval method that adopts both vertical and horizontal self-attentions to capture table structures and produces a joint representation for table rows and query tokens to predict the relevance score between the table and the query.

\item \Santos{}~\cite{khatiwada_santos_2023} (\emph{SIGMOD}'23), a table union search method that harnesses external knowledge bases (KBs) to identify columns and their binary relationships. 
It synthesizes KBs from the data lake (i.e., the table repository) to mitigate limited KB coverage.

\item \Starmie{}~\cite{fan_semantics-aware_2023} (\emph{VLDB}'23), a table union search method leveraging pre-trained language models. It employs contrastive learning to train column encoders, capturing contextual and semantic information in a fully unsupervised manner.

\item \Josie{}~\cite{zhu_josie_2019} (\emph{SIGMOD}'19), a classic table join search method that uses set similarity to identify joinable columns based on overlapping values. It employs an inverted index and cost models for efficient candidate filtering.

\item \DeepJoin{}~\cite{dong_deepjoin_2023} (\emph{VLDB}'23), a neural table join search method that fine-tunes pre-trained models like \textsc{DistilBERT}~\cite{Sanh2019DistilBERTAD} and \textsc{MPNet}~\cite{song2020mpnet}. It computes cosine similarity between column embeddings and indexes them using HNSW~\cite{malkov_efficient_2020} for efficient retrieval given the online query column embedding.

\end{enumerate}

\subsection{Overall Effectiveness Comparison (RQ1)}
\label{ssec:overall_comparison}

We test each dataset type using feasible methods capable of handling the corresponding query type, i.e., No.~1--2 on \dataset{}\textsf{-K}, No.~1--4 on \dataset{}\textsf{-U}, and No.~1--2 and 
 5--6 on \dataset{}\textsf{-J}.   

\smallskip
\noindent\textbf{\noindent\dataset{}\textsf{-K}} (Table~\ref{nl-only-result}): 
Both keyword-based methods show limited performance on \dataset{}\textsf{-K}, which contains long, sentence-based queries. Notably, for $k=10$, \Strubert{} performs worse, likely due to its use of cell-wise pooling, which is better suited for keyword-based queries. In contrast, \GTR{} converts entire tables into graphs, allowing it to better capture table context and perform relatively better on longer sentences.

\begin{table}[!htbp]
\caption{\GTR{} and \Strubert{} tested on \dataset{}\textsf{-K}.}
\label{nl-only-result}
\footnotesize
\begin{tabular}{lrrrr}
\toprule
\textbf{Methods}                 & $k$   & $P$      & $R$      & $\mathit{NDCG}$   \\ \midrule
\multirow{2}{*}{\GTR{}}      & 5  & 0.3155 & 0.5437 & 0.4577 \\
                          & 10 & 0.2155 & 0.7451 & 0.5441 \\\midrule
\multirow{2}{*}{\Strubert{}} & 5  & 0.1777 & 0.5569 & 0.4984 \\
                          & 10 & 0.0951 & 0.5952 & 0.5130  \\\bottomrule
\end{tabular}
\end{table}

\begin{table*}[ht]
\centering
\caption{The $\mathit{NDCG}@k$ results across condition categories in \dataset{}\textsf{-U}, with \GTR{} and \Starmie{} selected as the stronger competitors in their respective categories.}
\label{table:category}
\footnotesize
\renewcommand{\arraystretch}{0.9}
\begin{tabular}{lcccccccccc}
\toprule
\multirow{2}{*}{\textbf{Methods}} & \multirow{2}{*}{\textbf{$k$}} & \multicolumn{3}{c}{\textbf{Table-level}}                          & \multicolumn{5}{c}{\textbf{Column-level}}                              & \textbf{Mixed-mode} \\ 
\cmidrule(lr){3-5} \cmidrule(lr){6-10}
                                  &                               & \textbf{Topic} & \textbf{Size} & \textbf{Avg}         & \textbf{Categorical} & \textbf{String} & \textbf{Numerical} & \textbf{Date} & \textbf{Avg} & \textbf{Avg} \\ 
\midrule
\multirow{2}{*}{\GTR{}}           & 5                             & 0.9956         & 0.2054        & 0.6005               & 0.4678               & 0.1541          & 0.2763             & 0.3533        & 0.3129       & 0.1885       \\ 
                                  & 10                            & 0.9973         & 0.2582        & 0.6278               & 0.5146               & 0.3434          & 0.4183             & 0.4274        & 0.4259       & 0.3486       \\ 
\midrule
\multirow{2}{*}{\Starmie{}}       & 5                             & 0.9933         & 0.1949        & 0.5941               & 0.3667               & 0.3548          & 0.3562             & 0.2868        & 0.3411       & 0.1404       \\ 
                                  & 10                            & 0.9957         & 0.2764        & 0.6360               & 0.4844               & 0.5676          & 0.4717             & 0.5042        & 0.5070       & 0.1866       \\ 
\bottomrule
\end{tabular}
\end{table*}

\begin{table*}[ht]
\centering
\caption{Dataset scale vs efficiency and accuracy on \dataset{}\textsf{-U} and \dataset{}\textsf{-J}.
Efficiency (unit: \emph{second}) is evaluated in offline (e.g., embedding and index construction) and online (searching) stages.
}
\label{table:dataset-scale}
\resizebox{\textwidth}{!}{%
\renewcommand{\arraystretch}{0.95}
\begin{tabular}{@{}lrrrrrrrrrrrr@{}}
\toprule

\multirow{2}{*}{\dataset{}\textsf{-U}}          & \multicolumn{4}{c}{\textsf{Small} (16.7\%, $|\mathcal{T}|=1,620$)}                                & \multicolumn{4}{c}{\textsf{Medium} (50\%, $|\mathcal{T}|=5,006$)}                              & \multicolumn{4}{c}{\textsf{Large} (100\%, $|\mathcal{T}|=7,567$)}                              \\ \cmidrule(lr){2-5} \cmidrule(lr){6-9} \cmidrule(lr){10-13}
         & \multicolumn{1}{c}{Offline (s)} & \multicolumn{1}{c}{Online (s)} & \multicolumn{1}{c}{$\mathit{NDCG}@5$} & $\mathit{NDCG}@10$ & \multicolumn{1}{c}{Offline (s)} & \multicolumn{1}{c}{Online (s)} & \multicolumn{1}{c}{$\mathit{NDCG}@5$} & $\mathit{NDCG}@10$ & \multicolumn{1}{c}{Offline (s)} & \multicolumn{1}{c}{Online (s)} & \multicolumn{1}{c}{$\mathit{NDCG}@5$} & $\mathit{NDCG}@10$ \\ \midrule
\GTR{}      & 10,743.81            & 76.82              & 0.3960  & 0.4607  & 26,942.19            & 173.85            & 0.3605 & 0.453  & 53,095.33            & 752.39            & 0.3534 & 0.4491  \\
\Strubert{} & 376.02               & 172.44             & 0.6620  & 0.6899  & 1,230.01             & 297.59            & 0.569 & 0.5874  & 1,865.21             & 471.59            & 0.5139 & 0.5291  \\
\Santos{}   & 434.28               & 16.00              & 0.2874 & 0.3801  & 1,090.28             & 75.00             & 0.2536 & 0.3003  & 1,657.74             & 283.00            & 0.2516 & 0.3287  \\
\Starmie{}  & 10.15                & 56.19              & 0.3901 & 0.4816  & 27.48                & 311.85            & 0.3564  & 0.4271  & 31.92                & 1,573.65          & 0.3235 & 0.3892  \\ \midrule

 \multirow{2}{*}{\dataset{}\textsf{-J}}        & \multicolumn{4}{c}{\textsf{Small} (16.7\%, $|\mathcal{T}|=1,613$)}                                & \multicolumn{4}{c}{\textsf{Medium} (50\%, $|\mathcal{T}|=4,871$)}                              & \multicolumn{4}{c}{\textsf{Large} (100\%, $|\mathcal{T}|=9,347$)}                              \\ \cmidrule(lr){2-5} \cmidrule(lr){6-9} \cmidrule(lr){10-13}
         & \multicolumn{1}{c}{Offline (s)} & \multicolumn{1}{c}{Online (s)} & \multicolumn{1}{c}{$\mathit{NDCG}@5$} & $\mathit{NDCG}@10$ & \multicolumn{1}{c}{Offline (s)} & \multicolumn{1}{c}{Online (s)} & \multicolumn{1}{c}{$\mathit{NDCG}@5$} & $\mathit{NDCG}@10$ & \multicolumn{1}{c}{Offline (s)} & \multicolumn{1}{c}{Online (s)} & \multicolumn{1}{c}{$\mathit{NDCG}@5$} & $\mathit{NDCG}@10$ \\ \midrule
\GTR{}      & 8,953.15             & 188.38             & 0.3384 & 0.4106  & 30,802.16            & 182.06            & 0.3174  & 0.3916  & 61,163.23            & 357.17            & 0.3077 & 0.3852  \\
\Strubert{} & 334.55               & 123.99             & 0.4528 & 0.4650   & 3,109.74             & 310.22            & 0.458 & 0.4626  & 2,863.73             & 548.23            & 0.4358 & 0.4437  \\
\Josie{}    & 6.87                 & 0.33               & 0.4871 & 0.4758  & 55.51                & 11.77             & 0.4511 & 0.4479  & 73.88                & 45.86             & 0.4189 & 0.4023  \\
\DeepJoin{} & 404.93               & 17.29              & 0.3023 & 0.3879  & 1,305.00             & 201.15            & 0.2849 & 0.3627  & 2,502.25             & 204.20            & 0.2785 & 0.3631  \\ 
 \bottomrule
\end{tabular}%
}
\end{table*}

\begin{figure}[!htbp]
  \centering
  \includegraphics[width=\linewidth]{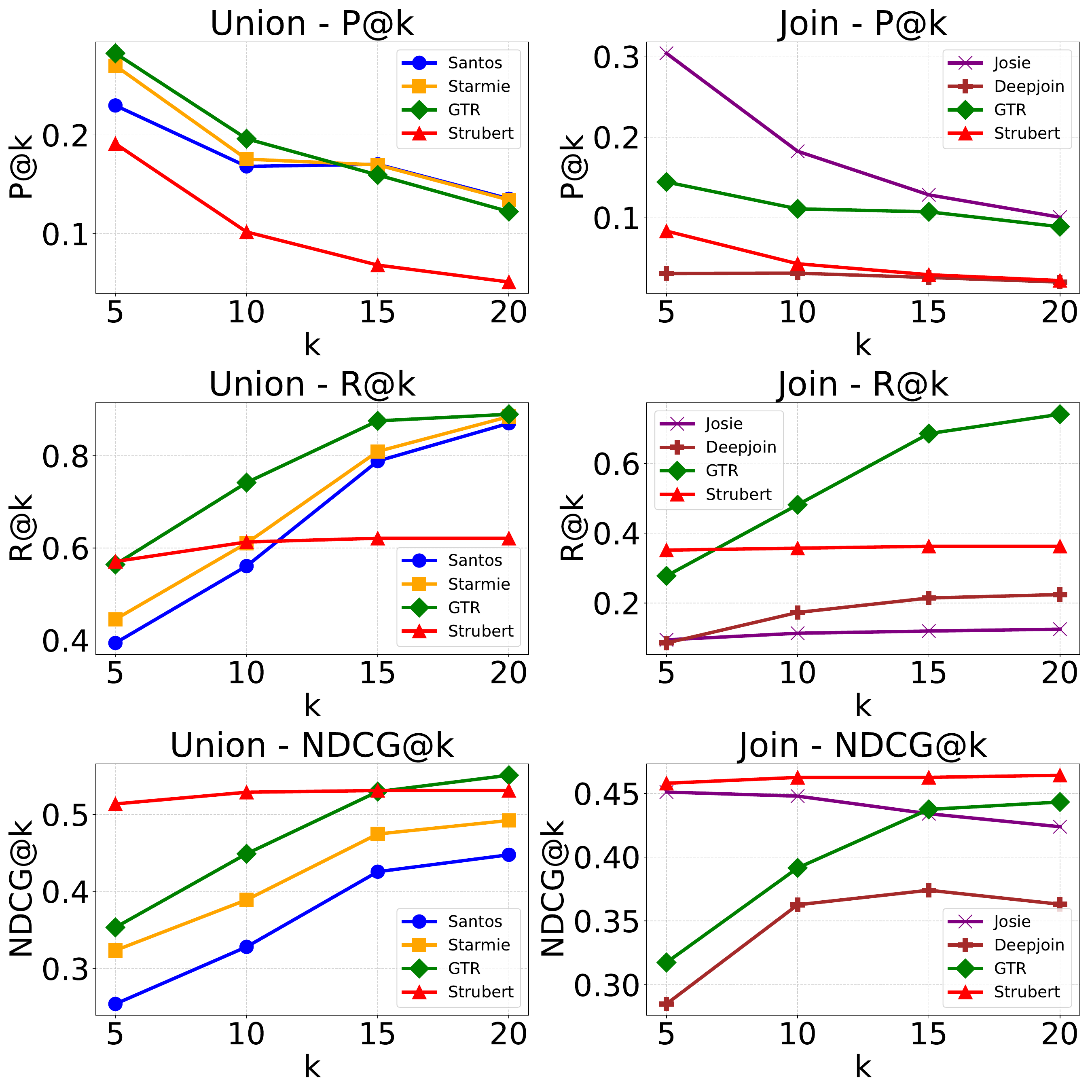}
  \caption{Comparisons of feasible methods on (left) \dataset{}\textsf{-U} and (right) \dataset{}\textsf{-J}.}
  \label{fig:overall_comparison}
\end{figure}

\smallskip
\noindent\textbf{\dataset{}\textsf{-U}} (Figure~\ref{fig:overall_comparison}(left)):
Overall, these four methods are not well-suited for NL-conditional table union search.
For table union search methods \Santos{} and \Starmie{}, their performance is suboptimal when $k \leq 15$, with a relatively high $R@20$ likely due to retrieving most unionable tables, including those meeting the conditions.
However, these two primarily focus on recalling tables related to the query table, regardless of whether they satisfy NL conditions.
As a result, they are unsuitable for scenarios requiring both query-table relevance and NL condition fulfillment. For example, a query table about students may lead to retrieving tables like ``student physical examination forms'', which fail to meet NL conditions, such as ``students with high grades''.
For keyword-based \Strubert{}, its performance declines in precision ($P$) when $k \geq 10$ without a notable increase in recall ($R$), indicating it retrieves very few relevant tables.
While \GTR{} performs slightly better, its sorting quality remains inadequate. Keyword-based methods often retrieve tables unrelated to the query table, making them ineffective for scenarios where users seek specific tables to enhance or supplement the original table.

\smallskip
\noindent\textbf{\dataset{}\textsf{-J}} (Figure~\ref{fig:overall_comparison}(right)):
Overall, none of these four methods are well-suited for joinable table search combined with NL conditions.
The query-table-based methods \Josie{} and \DeepJoin{} exhibit extremely low recall, indicating that they fail to identify relevant tables beyond the first few retrieved.
They are limited to retrieving joinable tables, which often do not satisfy the NL conditions, making them unsuitable for this task.
Keyword-based \Strubert{} struggles with retrieving joinable tables, performing poorly when $k \geq 10$. Similarly, \GTR{} performs poorly at $k=5$, suggesting that the initial retrieved tables are not joinable.

\subsection{Studies on Condition Categories (RQ2)}
\label{ssec:condition_categories}

Table~\ref{table:category} presents the $\mathit{NDCG}@k$ results across different condition categories, taking \dataset{}\textsf{-U} as an example. 
Overall, the performance of the tested methods varies significantly depending on the type of condition.
Most methods perform quite well on table topic queries, likely because topics are commonly used as keywords, and unionable or joinable tables often share the same topic.
However, these methods struggle with numerical and date columns. Existing techniques are primarily designed to retrieve related tables but often overlook these specialized column types.
Mixed-mode conditions present an even greater challenge due to their inherent complexity.
As shown in the statistics in Table~\ref{table:statistic}, our \dataset{} places greater emphasis on the more challenging conditions, such as column-level and mixed-mode, providing ample room for improvement and testing the capabilities of future approaches.

\subsection{Studies on Dataset Composition (RQ3)}
\label{ssec:benchmark_composition}

Given the configurable nature of \dataset{}, we conduct experiments to analyze its composition, focusing on the following factors.

\begin{figure}[!htbp]
  \centering
  \includegraphics[width=0.85\linewidth]{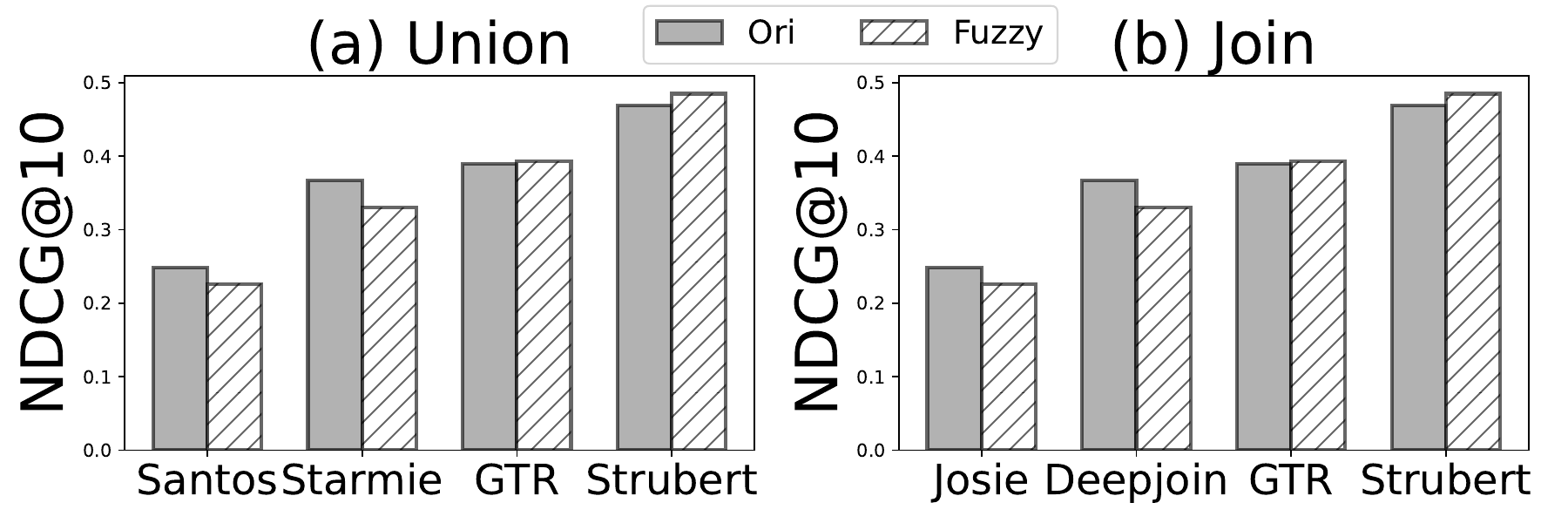}
  \caption{Comparisons of feasible methods on (a) \dataset{}\textsf{-U-fz} and (b) \dataset{}\textsf{-J-fz} for fuzzy queries.}
  \label{fig:fuzzy}
\end{figure}

\noindent\textbf{Fuzzy Queries} (Figure~\ref{fig:fuzzy}):
Query-table-based methods are more affected by the incorporation of fuzzy queries generated by semantic augmentation, likely due to their reliance on table similarity.
If semantic similarity is not preserved via transformation, these methods often fail.
This highlights that capturing nuanced table semantics remains a challenge for query-table-based approaches.
In contrast, keyword-based methods are largely unaffected, as they rely on keyword-to-table similarity.
Changes in columns have minimal impact on keywords, and language models used in these methods are more sensitive to similar terms.

\begin{figure}[!htbp]
  \centering
  \includegraphics[width=0.9\linewidth]{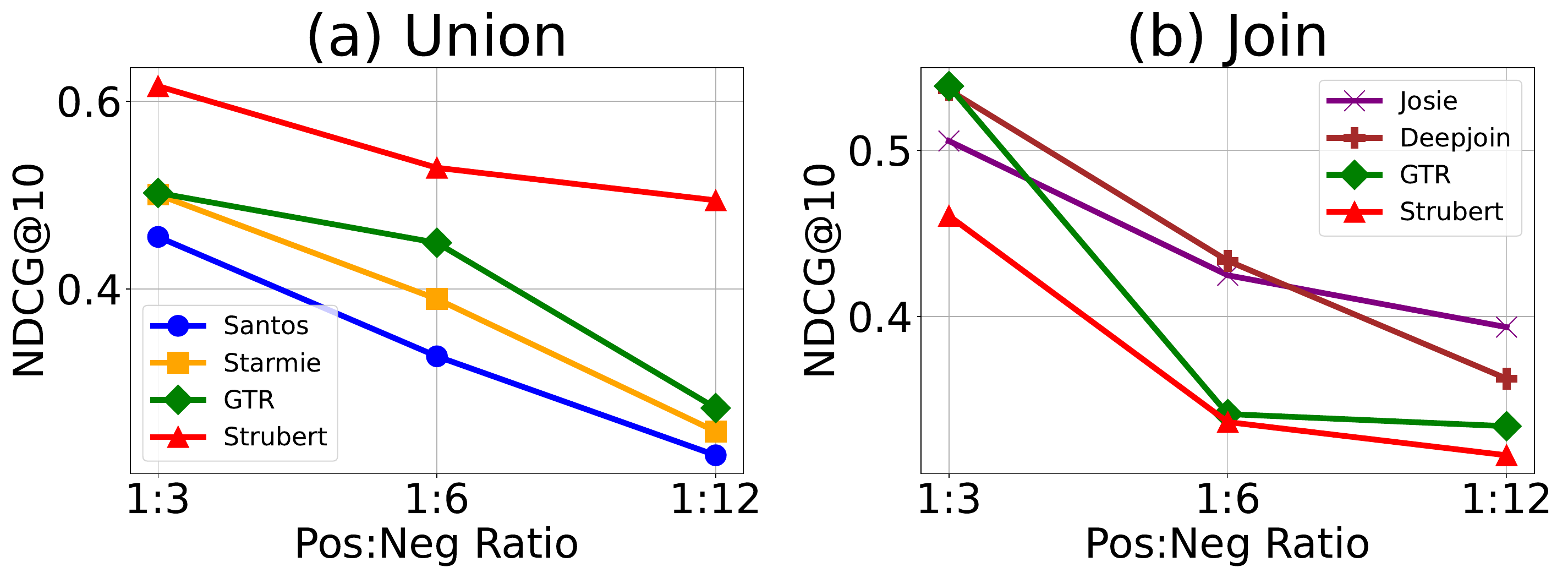}
  \caption{$\mathit{NDCG}@k$ vs positive-to-negative ratio.}
  \Description{}
  \label{fig:neg}
\end{figure}

\noindent\textbf{Positive-to-Negative Sample Ratio} (Figure~\ref{fig:neg}):
As the number of negative samples increases, accuracy declines clearly for all methods for both tasks. This is due to the introduction of more misleading yet similar tables, increasing the dataset's difficulty. This demonstrates the high usability of \dataset{}.

\noindent\textbf{Dataset Scale} (Table~\ref{table:dataset-scale}):
We derive \dataset{} of \textsf{Small}, \textsf{Medium}, and \textsf{Large} sizes (see Section~\ref{statistics}) for efficiency analysis.
Methods employing specialized indexing structures, such as inverted indexes (\Josie{}) or HNSW (\Starmie{} and \DeepJoin{}), demonstrate higher efficiency. In contrast, \GTR{}, \Strubert{}, and \Santos{} require clearly more time without offering noticeable accuracy advantages.

\smallskip
\noindent\textbf{Summary}. 
Effective table semantics capturing is a promising direction for \task{}. 
As datasets become more challenging, most existing methods fall short, underscoring the need for more robust approaches. In addition, it is critical to strike a balance between efficiency and accuracy in the solution toward \task{}.

\section{Conclusion}

We introduce a novel table discovery scenario, \task{}, along with a comprehensive dataset and its construction framework. 
The following resources are available in our GitHub repository:
\begin{itemize}[leftmargin=*]
\item \textbf{Scripts}, we provide scripts for dataset construction, allowing users to adjust hyperparameters and create their own 
testing datasets to suit specific needs.

\item \textbf{Dataset}, we release \dataset{} of 22,080 tables, 627 queries, and 21,200 corresponding ground truth annotations.

\item \textbf{Baselines}, we publish implementations of the baselines used in our study to facilitate result reproduction by others.
\end{itemize}

Our proposed \task{} task introduces a new paradigm for table discovery, aligning with the growing emergence of AI-powered tools for data science and information retrieval. 
As the first large-scale and diverse dataset for \task{}, \dataset{} offers a foundational resource for advancing future research in this field.

\section*{Acknowledgments}
This research was funded by the Major Research Program of the Zhejiang Provincial Natural Science Foundation (Grant No.~LD24F020015), the Pioneer R\&D Program of Zhejiang (Grant No.~2024C01021), NSFC Grant No.~U24A201401, and Zhejiang Province ``Leading Talent of Technological Innovation Program'' (No. 2023R5214).

\balance
\clearpage
\bibliographystyle{ACM-Reference-Format}
\bibliography{main}


\begin{thebibliography}{41}


\ifx \showCODEN    \undefined \def \showCODEN     #1{\unskip}     \fi
\ifx \showISBNx    \undefined \def \showISBNx     #1{\unskip}     \fi
\ifx \showISBNxiii \undefined \def \showISBNxiii  #1{\unskip}     \fi
\ifx \showISSN     \undefined \def \showISSN      #1{\unskip}     \fi
\ifx \showLCCN     \undefined \def \showLCCN      #1{\unskip}     \fi
\ifx \shownote     \undefined \def \shownote      #1{#1}          \fi
\ifx \showarticletitle \undefined \def \showarticletitle #1{#1}   \fi
\ifx \showURL      \undefined \def \showURL       {\relax}        \fi
\providecommand\bibfield[2]{#2}
\providecommand\bibinfo[2]{#2}
\providecommand\natexlab[1]{#1}
\providecommand\showeprint[2][]{arXiv:#2}

\bibitem[Cop({[n.\,d.]})]%
        {Copilot_github}
 \bibinfo{year}{[n.\,d.]}\natexlab{}.
\newblock \bibinfo{title}{Copilot}.
\newblock
\urldef\tempurl%
\url{https://github.com/features/copilot}
\showURL{%
\tempurl}


\bibitem[ope({[n.\,d.]})]%
        {opendata}
 \bibinfo{year}{[n.\,d.]}\natexlab{}.
\newblock \bibinfo{title}{OpenData}.
\newblock
\urldef\tempurl%
\url{https://open.canada.ca/}
\showURL{%
\tempurl}


\bibitem[Bogatu et~al\mbox{.}(2020)]%
        {bogatu_dataset_2020}
\bibfield{author}{\bibinfo{person}{Alex Bogatu}, \bibinfo{person}{Alvaro A.~A. Fernandes}, \bibinfo{person}{Norman~W. Paton}, {and} \bibinfo{person}{Nikolaos Konstantinou}.} \bibinfo{year}{2020}\natexlab{}.
\newblock \showarticletitle{Dataset {Discovery} in {Data} {Lakes}}. In \bibinfo{booktitle}{\emph{Proc. 36th {IEEE} {ICDE}}}. \bibinfo{pages}{709--720}.
\newblock
\href{https://doi.org/10.1109/ICDE48307.2020.00067}{doi:\nolinkurl{10.1109/ICDE48307.2020.00067}}


\bibitem[Cafarella et~al\mbox{.}(2009)]%
        {cafarella2009data}
\bibfield{author}{\bibinfo{person}{Michael~J Cafarella}, \bibinfo{person}{Alon Halevy}, {and} \bibinfo{person}{Nodira Khoussainova}.} \bibinfo{year}{2009}\natexlab{}.
\newblock \showarticletitle{Data integration for the relational web}.
\newblock \bibinfo{journal}{\emph{Proc. VLDB Endow.}} \bibinfo{volume}{2}, \bibinfo{number}{1} (\bibinfo{year}{2009}), \bibinfo{pages}{1090--1101}.
\newblock


\bibitem[Chai et~al\mbox{.}(2024)]%
        {lakebench2024chai}
\bibfield{author}{\bibinfo{person}{Chengliang Chai}, \bibinfo{person}{Yuhao Deng}, \bibinfo{person}{Yutong Zhan}, \bibinfo{person}{Ziqi Cao}, \bibinfo{person}{Yuanfang Zhang}, \bibinfo{person}{Lei Cao}, \bibinfo{person}{Yuping Wang}, \bibinfo{person}{Zhiwei Zhang}, \bibinfo{person}{Ye Yuan}, \bibinfo{person}{Guoren Wang}, {and} \bibinfo{person}{Nan Tang}.} \bibinfo{year}{2024}\natexlab{}.
\newblock \showarticletitle{LakeCompass: An End-to-End System for Data Maintenance, Search and Analysis in Data Lakes}.
\newblock \bibinfo{journal}{\emph{Proc. VLDB Endow.}} \bibinfo{volume}{17}, \bibinfo{number}{12} (\bibinfo{year}{2024}), \bibinfo{pages}{4381–4384}.
\newblock
\href{https://doi.org/10.14778/3685800.3685880}{doi:\nolinkurl{10.14778/3685800.3685880}}


\bibitem[Chen et~al\mbox{.}(2024)]%
        {tablerag_2024_Chen}
\bibfield{author}{\bibinfo{person}{Si-An Chen}, \bibinfo{person}{Lesly Miculicich}, \bibinfo{person}{Julian Eisenschlos}, \bibinfo{person}{Zifeng Wang}, \bibinfo{person}{Zilong Wang}, \bibinfo{person}{Yanfei Chen}, \bibinfo{person}{YASUHISA FUJII}, \bibinfo{person}{Hsuan-Tien Lin}, \bibinfo{person}{Chen-Yu Lee}, {and} \bibinfo{person}{Tomas Pfister}.} \bibinfo{year}{2024}\natexlab{}.
\newblock \showarticletitle{TableRAG: Million-Token Table Understanding with Language Models}. In \bibinfo{booktitle}{\emph{NeurIPS}}, Vol.~\bibinfo{volume}{37}. \bibinfo{pages}{74899--74921}.
\newblock


\bibitem[Chen et~al\mbox{.}(2020)]%
        {Table2020Zhiyu}
\bibfield{author}{\bibinfo{person}{Zhiyu Chen}, \bibinfo{person}{Mohamed Trabelsi}, \bibinfo{person}{Jeff Heflin}, \bibinfo{person}{Yinan Xu}, {and} \bibinfo{person}{Brian~D. Davison}.} \bibinfo{year}{2020}\natexlab{}.
\newblock \showarticletitle{Table Search Using a Deep Contextualized Language Model}. In \bibinfo{booktitle}{\emph{Proc. SIGIR. ACM}}. \bibinfo{pages}{589–598}.
\newblock
\href{https://doi.org/10.1145/3397271.3401044}{doi:\nolinkurl{10.1145/3397271.3401044}}


\bibitem[Chepurko et~al\mbox{.}(2020)]%
        {chepurko_arda_2020}
\bibfield{author}{\bibinfo{person}{Nadiia Chepurko}, \bibinfo{person}{Ryan Marcus}, \bibinfo{person}{Emanuel Zgraggen}, \bibinfo{person}{Raul~Castro Fernandez}, \bibinfo{person}{Tim Kraska}, {and} \bibinfo{person}{David Karger}.} \bibinfo{year}{2020}\natexlab{}.
\newblock \showarticletitle{{ARDA}: automatic relational data augmentation for machine learning}.
\newblock \bibinfo{journal}{\emph{Proc. VLDB Endow.}} \bibinfo{volume}{13}, \bibinfo{number}{9} (\bibinfo{year}{2020}), \bibinfo{pages}{1373--1387}.
\newblock
\href{https://doi.org/10.14778/3397230.3397235}{doi:\nolinkurl{10.14778/3397230.3397235}}


\bibitem[Cui et~al\mbox{.}(2024)]%
        {Cui2024TabularDA}
\bibfield{author}{\bibinfo{person}{Lingxi Cui}, \bibinfo{person}{Huan Li}, \bibinfo{person}{Ke Chen}, \bibinfo{person}{Lidan Shou}, {and} \bibinfo{person}{Gang Chen}.} \bibinfo{year}{2024}\natexlab{}.
\newblock \showarticletitle{Tabular Data Augmentation for Machine Learning: Progress and Prospects of Embracing Generative AI}.
\newblock \bibinfo{journal}{\emph{ArXiv}}  \bibinfo{volume}{abs/2407.21523} (\bibinfo{year}{2024}).
\newblock
\href{https://doi.org/10.48550/arXiv.2407.21523}{doi:\nolinkurl{10.48550/arXiv.2407.21523}}


\bibitem[Deng et~al\mbox{.}(2020)]%
        {deng_turl_2020}
\bibfield{author}{\bibinfo{person}{Xiang Deng}, \bibinfo{person}{Huan Sun}, \bibinfo{person}{Alyssa Lees}, \bibinfo{person}{You Wu}, {and} \bibinfo{person}{Cong Yu}.} \bibinfo{year}{2020}\natexlab{}.
\newblock \showarticletitle{{TURL}: table understanding through representation learning}.
\newblock \bibinfo{journal}{\emph{Proc. VLDB Endow.}} \bibinfo{volume}{14}, \bibinfo{number}{3} (\bibinfo{year}{2020}), \bibinfo{pages}{307--319}.
\newblock
\href{https://doi.org/10.14778/3430915.3430921}{doi:\nolinkurl{10.14778/3430915.3430921}}


\bibitem[Deng et~al\mbox{.}(2024)]%
        {deng2024lakebench}
\bibfield{author}{\bibinfo{person}{Yuhao Deng}, \bibinfo{person}{Chengliang Chai}, \bibinfo{person}{Lei Cao}, \bibinfo{person}{Qin Yuan}, \bibinfo{person}{Siyuan Chen}, \bibinfo{person}{Yanrui Yu}, \bibinfo{person}{Zhaoze Sun}, \bibinfo{person}{Junyi Wang}, \bibinfo{person}{Jiajun Li}, \bibinfo{person}{Ziqi Cao}, {et~al\mbox{.}}} \bibinfo{year}{2024}\natexlab{}.
\newblock \showarticletitle{LakeBench: A Benchmark for Discovering Joinable and Unionable Tables in Data Lakes}.
\newblock \bibinfo{journal}{\emph{Proc. VLDB Endow.}} \bibinfo{volume}{17}, \bibinfo{number}{8} (\bibinfo{year}{2024}), \bibinfo{pages}{1925--1938}.
\newblock


\bibitem[Dong and Oyamada(2022)]%
        {Table2022dong}
\bibfield{author}{\bibinfo{person}{Yuyang Dong} {and} \bibinfo{person}{Masafumi Oyamada}.} \bibinfo{year}{2022}\natexlab{}.
\newblock \showarticletitle{Table Enrichment System for Machine Learning}. In \bibinfo{booktitle}{\emph{Proc. 45th SIGIR. ACM.}} \bibinfo{pages}{3267–3271}.
\newblock
\href{https://doi.org/10.1145/3477495.3531678}{doi:\nolinkurl{10.1145/3477495.3531678}}


\bibitem[Dong et~al\mbox{.}(2021)]%
        {dong_efficient_2021}
\bibfield{author}{\bibinfo{person}{Yuyang Dong}, \bibinfo{person}{Kunihiro Takeoka}, \bibinfo{person}{Chuan Xiao}, {and} \bibinfo{person}{Masafumi Oyamada}.} \bibinfo{year}{2021}\natexlab{}.
\newblock \showarticletitle{Efficient {Joinable} {Table} {Discovery} in {Data} {Lakes}: {A} {High}-{Dimensional} {Similarity}-{Based} {Approach}}. In \bibinfo{booktitle}{\emph{Proc. 37th {IEEE} {ICDE}}}. \bibinfo{pages}{456--467}.
\newblock
\href{https://doi.org/10.1109/ICDE51399.2021.00046}{doi:\nolinkurl{10.1109/ICDE51399.2021.00046}}


\bibitem[Dong et~al\mbox{.}(2023)]%
        {dong_deepjoin_2023}
\bibfield{author}{\bibinfo{person}{Yuyang Dong}, \bibinfo{person}{Chuan Xiao}, \bibinfo{person}{Takuma Nozawa}, \bibinfo{person}{Masafumi Enomoto}, {and} \bibinfo{person}{Masafumi Oyamada}.} \bibinfo{year}{2023}\natexlab{}.
\newblock \showarticletitle{{DeepJoin}: {Joinable} {Table} {Discovery} with {Pre}-{Trained} {Language} {Models}}.
\newblock \bibinfo{journal}{\emph{Proc. VLDB Endow.}} \bibinfo{volume}{16}, \bibinfo{number}{10} (\bibinfo{year}{2023}), \bibinfo{pages}{2458--2470}.
\newblock
\href{https://doi.org/10.14778/3603581.3603587}{doi:\nolinkurl{10.14778/3603581.3603587}}


\bibitem[Fan et~al\mbox{.}(2023a)]%
        {fan_table_2023}
\bibfield{author}{\bibinfo{person}{Grace Fan}, \bibinfo{person}{Jin Wang}, \bibinfo{person}{Yuliang Li}, {and} \bibinfo{person}{Renée~J. Miller}.} \bibinfo{year}{2023}\natexlab{a}.
\newblock \showarticletitle{Table {Discovery} in {Data} {Lakes}: {State}-of-the-art and {Future} {Directions}}. In \bibinfo{booktitle}{\emph{Companion of SIGMOD}}. \bibinfo{pages}{69--75}.
\newblock
\href{https://doi.org/10.1145/3555041.3589409}{doi:\nolinkurl{10.1145/3555041.3589409}}


\bibitem[Fan et~al\mbox{.}(2023b)]%
        {fan_semantics-aware_2023}
\bibfield{author}{\bibinfo{person}{Grace Fan}, \bibinfo{person}{Jin Wang}, \bibinfo{person}{Yuliang Li}, \bibinfo{person}{Dan Zhang}, {and} \bibinfo{person}{Ren{\'{e}}e~J. Miller}.} \bibinfo{year}{2023}\natexlab{b}.
\newblock \showarticletitle{Semantics-aware Dataset Discovery from Data Lakes with Contextualized Column-based Representation Learning}.
\newblock \bibinfo{journal}{\emph{Proc. {VLDB} Endow.}} \bibinfo{volume}{16}, \bibinfo{number}{7} (\bibinfo{year}{2023}), \bibinfo{pages}{1726--1739}.
\newblock
\href{https://doi.org/10.14778/3587136.3587146}{doi:\nolinkurl{10.14778/3587136.3587146}}


\bibitem[Hu et~al\mbox{.}(2023)]%
        {hu_automatic_2023}
\bibfield{author}{\bibinfo{person}{Xuming Hu}, \bibinfo{person}{Shen Wang}, \bibinfo{person}{Xiao Qin}, \bibinfo{person}{Chuan Lei}, \bibinfo{person}{Zhengyuan Shen}, \bibinfo{person}{Christos Faloutsos}, \bibinfo{person}{Asterios Katsifodimos}, \bibinfo{person}{George Karypis}, \bibinfo{person}{Lijie Wen}, {and} \bibinfo{person}{Philip~S. Yu}.} \bibinfo{year}{2023}\natexlab{}.
\newblock \showarticletitle{Automatic {Table} {Union} {Search} with {Tabular} {Representation} {Learning}}. In \bibinfo{booktitle}{\emph{Findings of the {Association} for {Computational} {Linguistics}: {ACL} 2023}}. \bibinfo{pages}{3786--3800}.
\newblock
\href{https://doi.org/10.18653/v1/2023.findings-acl.233}{doi:\nolinkurl{10.18653/v1/2023.findings-acl.233}}


\bibitem[Hulsebos et~al\mbox{.}(2023)]%
        {hulsebos2023gittables}
\bibfield{author}{\bibinfo{person}{Madelon Hulsebos}, \bibinfo{person}{\c{C}agatay Demiralp}, {and} \bibinfo{person}{Paul Groth}.} \bibinfo{year}{2023}\natexlab{}.
\newblock \showarticletitle{GitTables: A Large-Scale Corpus of Relational Tables}.
\newblock \bibinfo{journal}{\emph{Proc. ACM Manag. Data}} \bibinfo{volume}{1}, \bibinfo{number}{1} (\bibinfo{year}{2023}), \bibinfo{pages}{1--17}.
\newblock
\href{https://doi.org/10.1145/3588710}{doi:\nolinkurl{10.1145/3588710}}


\bibitem[Khatiwada et~al\mbox{.}(2023)]%
        {khatiwada_santos_2023}
\bibfield{author}{\bibinfo{person}{Aamod Khatiwada}, \bibinfo{person}{Grace Fan}, \bibinfo{person}{Roee Shraga}, \bibinfo{person}{Zixuan Chen}, \bibinfo{person}{Wolfgang Gatterbauer}, \bibinfo{person}{Renée~J. Miller}, {and} \bibinfo{person}{Mirek Riedewald}.} \bibinfo{year}{2023}\natexlab{}.
\newblock \showarticletitle{{SANTOS}: {Relationship}-based {Semantic} {Table} {Union} {Search}}.
\newblock \bibinfo{journal}{\emph{Proc. ACM Manag. Data}} \bibinfo{volume}{1}, \bibinfo{number}{1} (\bibinfo{year}{2023}), \bibinfo{pages}{1--25}.
\newblock
\href{https://doi.org/10.1145/3588689}{doi:\nolinkurl{10.1145/3588689}}


\bibitem[Leventidis et~al\mbox{.}(2024)]%
        {Leventidis2024ALS}
\bibfield{author}{\bibinfo{person}{Aristotelis Leventidis}, \bibinfo{person}{Martin~Pek{\'a}r Christensen}, \bibinfo{person}{Matteo Lissandrini}, \bibinfo{person}{Laura~Di Rocco}, \bibinfo{person}{Katja Hose}, {and} \bibinfo{person}{Ren{\'e}e~J. Miller}.} \bibinfo{year}{2024}\natexlab{}.
\newblock \showarticletitle{A Large Scale Test Corpus for Semantic Table Search}. In \bibinfo{booktitle}{\emph{Proc. 47th SIGIR. ACM.}}
\newblock
\urldef\tempurl%
\url{https://api.semanticscholar.org/CorpusID:271114571}
\showURL{%
\tempurl}


\bibitem[Li et~al\mbox{.}(2024)]%
        {li2024can}
\bibfield{author}{\bibinfo{person}{Jinyang Li}, \bibinfo{person}{Binyuan Hui}, \bibinfo{person}{Ge Qu}, \bibinfo{person}{Jiaxi Yang}, \bibinfo{person}{Binhua Li}, \bibinfo{person}{Bowen Li}, \bibinfo{person}{Bailin Wang}, \bibinfo{person}{Bowen Qin}, \bibinfo{person}{Ruiying Geng}, \bibinfo{person}{Nan Huo}, {et~al\mbox{.}}} \bibinfo{year}{2024}\natexlab{}.
\newblock \showarticletitle{Can llm already serve as a database interface? a big bench for large-scale database grounded text-to-sqls}.
\newblock \bibinfo{journal}{\emph{NeurIPS}}  \bibinfo{volume}{36}.
\newblock


\bibitem[Liu et~al\mbox{.}(2022)]%
        {liu_feature_2022}
\bibfield{author}{\bibinfo{person}{Jiabin Liu}, \bibinfo{person}{Chengliang Chai}, \bibinfo{person}{Yuyu Luo}, \bibinfo{person}{Yin Lou}, \bibinfo{person}{Jianhua Feng}, {and} \bibinfo{person}{Nan Tang}.} \bibinfo{year}{2022}\natexlab{}.
\newblock \showarticletitle{Feature {Augmentation} with {Reinforcement} {Learning}}. In \bibinfo{booktitle}{\emph{Proc. 38th {IEEE} {ICDE}}}. \bibinfo{pages}{3360--3372}.
\newblock
\href{https://doi.org/10.1109/ICDE53745.2022.00317}{doi:\nolinkurl{10.1109/ICDE53745.2022.00317}}


\bibitem[Malkov and Yashunin(2020)]%
        {malkov_efficient_2020}
\bibfield{author}{\bibinfo{person}{Yu~A. Malkov} {and} \bibinfo{person}{D.~A. Yashunin}.} \bibinfo{year}{2020}\natexlab{}.
\newblock \showarticletitle{Efficient and {Robust} {Approximate} {Nearest} {Neighbor} {Search} {Using} {Hierarchical} {Navigable} {Small} {World} {Graphs}}.
\newblock \bibinfo{journal}{\emph{IEEE Trans. Pattern Anal. Mach. Intell.}} \bibinfo{volume}{42}, \bibinfo{number}{4} (\bibinfo{year}{2020}), \bibinfo{pages}{824--836}.
\newblock
\href{https://doi.org/10.1109/TPAMI.2018.2889473}{doi:\nolinkurl{10.1109/TPAMI.2018.2889473}}


\bibitem[Nargesian et~al\mbox{.}(2018)]%
        {nargesian_table_2018}
\bibfield{author}{\bibinfo{person}{Fatemeh Nargesian}, \bibinfo{person}{Erkang Zhu}, \bibinfo{person}{Ken~Q. Pu}, {and} \bibinfo{person}{Renée~J. Miller}.} \bibinfo{year}{2018}\natexlab{}.
\newblock \showarticletitle{Table union search on open data}.
\newblock \bibinfo{journal}{\emph{Proc. VLDB Endow.}} \bibinfo{volume}{11}, \bibinfo{number}{7} (\bibinfo{year}{2018}), \bibinfo{pages}{813--825}.
\newblock
\href{https://doi.org/10.14778/3192965.3192973}{doi:\nolinkurl{10.14778/3192965.3192973}}


\bibitem[Paton et~al\mbox{.}(2024)]%
        {paton_dataset_2024}
\bibfield{author}{\bibinfo{person}{Norman~W. Paton}, \bibinfo{person}{Jiaoyan Chen}, {and} \bibinfo{person}{Zhenyu Wu}.} \bibinfo{year}{2024}\natexlab{}.
\newblock \showarticletitle{Dataset {Discovery} and {Exploration}: {A} {Survey}}.
\newblock \bibinfo{journal}{\emph{ACM Comput. Surv.}} \bibinfo{volume}{56}, \bibinfo{number}{4} (\bibinfo{year}{2024}), \bibinfo{pages}{1--37}.
\newblock
\href{https://doi.org/10.1145/3626521}{doi:\nolinkurl{10.1145/3626521}}


\bibitem[Petroni et~al\mbox{.}(2024)]%
        {IR4RAG2024}
\bibfield{author}{\bibinfo{person}{Fabio Petroni}, \bibinfo{person}{Federico Siciliano}, \bibinfo{person}{Fabrizio Silvestri}, {and} \bibinfo{person}{Giovanni Trappolini}.} \bibinfo{year}{2024}\natexlab{}.
\newblock \showarticletitle{IR-RAG @ SIGIR24: Information Retrieval's Role in RAG Systems}. In \bibinfo{booktitle}{\emph{Proc. 47th SIGIR. ACM.}} \bibinfo{pages}{3036–3039}.
\newblock
\href{https://doi.org/10.1145/3626772.3657984}{doi:\nolinkurl{10.1145/3626772.3657984}}


\bibitem[Sanh et~al\mbox{.}(2019)]%
        {Sanh2019DistilBERTAD}
\bibfield{author}{\bibinfo{person}{Victor Sanh}, \bibinfo{person}{Lysandre Debut}, \bibinfo{person}{Julien Chaumond}, {and} \bibinfo{person}{Thomas Wolf}.} \bibinfo{year}{2019}\natexlab{}.
\newblock \showarticletitle{DistilBERT, a distilled version of BERT: smaller, faster, cheaper and lighter}.
\newblock \bibinfo{journal}{\emph{ArXiv}}  \bibinfo{volume}{abs/1910.01108} (\bibinfo{year}{2019}).
\newblock
\urldef\tempurl%
\url{https://api.semanticscholar.org/CorpusID:203626972}
\showURL{%
\tempurl}


\bibitem[Santos et~al\mbox{.}(2022)]%
        {santos_sketch-based_2022}
\bibfield{author}{\bibinfo{person}{Aecio Santos}, \bibinfo{person}{Aline Bessa}, \bibinfo{person}{Christopher Musco}, {and} \bibinfo{person}{Juliana Freire}.} \bibinfo{year}{2022}\natexlab{}.
\newblock \showarticletitle{A {Sketch}-based {Index} for {Correlated} {Dataset} {Search}}. In \bibinfo{booktitle}{\emph{Proc. 38th {IEEE} {ICDE}}}. \bibinfo{pages}{2928--2941}.
\newblock
\href{https://doi.org/10.1109/ICDE53745.2022.00264}{doi:\nolinkurl{10.1109/ICDE53745.2022.00264}}


\bibitem[Shraga et~al\mbox{.}(2020)]%
        {Web2020Shraga}
\bibfield{author}{\bibinfo{person}{Roee Shraga}, \bibinfo{person}{Haggai Roitman}, \bibinfo{person}{Guy Feigenblat}, {and} \bibinfo{person}{Mustafa Cannim}.} \bibinfo{year}{2020}\natexlab{}.
\newblock \showarticletitle{Web Table Retrieval using Multimodal Deep Learning}. In \bibinfo{booktitle}{\emph{Proc. 43rd SIGIR. ACM}}. \bibinfo{pages}{1399–1408}.
\newblock
\href{https://doi.org/10.1145/3397271.3401120}{doi:\nolinkurl{10.1145/3397271.3401120}}


\bibitem[Song et~al\mbox{.}(2020)]%
        {song2020mpnet}
\bibfield{author}{\bibinfo{person}{Kaitao Song}, \bibinfo{person}{Xu Tan}, \bibinfo{person}{Tao Qin}, \bibinfo{person}{Jianfeng Lu}, {and} \bibinfo{person}{Tie-Yan Liu}.} \bibinfo{year}{2020}\natexlab{}.
\newblock \showarticletitle{Mpnet: Masked and permuted pre-training for language understanding}.
\newblock \bibinfo{journal}{\emph{NeurIPS}}  \bibinfo{volume}{33}, \bibinfo{pages}{16857--16867}.
\newblock


\bibitem[Trabelsi et~al\mbox{.}(2022)]%
        {trabelsi_strubert_2022}
\bibfield{author}{\bibinfo{person}{Mohamed Trabelsi}, \bibinfo{person}{Zhiyu Chen}, \bibinfo{person}{Shuo Zhang}, \bibinfo{person}{Brian~D. Davison}, {and} \bibinfo{person}{Jeff Heflin}.} \bibinfo{year}{2022}\natexlab{}.
\newblock \showarticletitle{{StruBERT}: {Structure}-aware {BERT} for {Table} {Search} and {Matching}}. In \bibinfo{booktitle}{\emph{Proceedings of the {ACM} {Web} {Conference} 2022}}. \bibinfo{pages}{442--451}.
\newblock
\href{https://doi.org/10.1145/3485447.3511972}{doi:\nolinkurl{10.1145/3485447.3511972}}


\bibitem[Venetis et~al\mbox{.}(2011)]%
        {venetis_recovering_2011}
\bibfield{author}{\bibinfo{person}{Petros Venetis}, \bibinfo{person}{Alon Halevy}, \bibinfo{person}{Jayant Madhavan}, \bibinfo{person}{Marius Paşca}, \bibinfo{person}{Warren Shen}, \bibinfo{person}{Fei Wu}, \bibinfo{person}{Gengxin Miao}, {and} \bibinfo{person}{Chung Wu}.} \bibinfo{year}{2011}\natexlab{}.
\newblock \showarticletitle{Recovering semantics of tables on the web}.
\newblock \bibinfo{journal}{\emph{Proc. VLDB Endow.}} \bibinfo{volume}{4}, \bibinfo{number}{9} (\bibinfo{year}{2011}), \bibinfo{pages}{528--538}.
\newblock
\href{https://doi.org/10.14778/2002938.2002939}{doi:\nolinkurl{10.14778/2002938.2002939}}


\bibitem[Wang et~al\mbox{.}(2021)]%
        {wang_retrieving_2021}
\bibfield{author}{\bibinfo{person}{Fei Wang}, \bibinfo{person}{Kexuan Sun}, \bibinfo{person}{Muhao Chen}, \bibinfo{person}{Jay Pujara}, {and} \bibinfo{person}{Pedro Szekely}.} \bibinfo{year}{2021}\natexlab{}.
\newblock \showarticletitle{Retrieving {Complex} {Tables} with {Multi}-{Granular} {Graph} {Representation} {Learning}}. In \bibinfo{booktitle}{\emph{Proc. 44th SIGIR. ACM}}. \bibinfo{pages}{1472--1482}.
\newblock
\href{https://doi.org/10.1145/3404835.3462909}{doi:\nolinkurl{10.1145/3404835.3462909}}


\bibitem[Wang and Fernandez(2023)]%
        {wang_solo_2023}
\bibfield{author}{\bibinfo{person}{Qiming Wang} {and} \bibinfo{person}{Raul~Castro Fernandez}.} \bibinfo{year}{2023}\natexlab{}.
\newblock \showarticletitle{Solo: Data Discovery Using Natural Language Questions Via {A} Self-Supervised Approach}.
\newblock \bibinfo{journal}{\emph{Proc. {ACM} Manag. Data}} \bibinfo{volume}{1}, \bibinfo{number}{4} (\bibinfo{year}{2023}), \bibinfo{pages}{262:1--262:27}.
\newblock
\href{https://doi.org/10.1145/3626756}{doi:\nolinkurl{10.1145/3626756}}


\bibitem[Yakout et~al\mbox{.}(2012)]%
        {yakout_infogather_2012}
\bibfield{author}{\bibinfo{person}{Mohamed Yakout}, \bibinfo{person}{Kris Ganjam}, \bibinfo{person}{Kaushik Chakrabarti}, {and} \bibinfo{person}{Surajit Chaudhuri}.} \bibinfo{year}{2012}\natexlab{}.
\newblock \showarticletitle{{InfoGather}: entity augmentation and attribute discovery by holistic matching with web tables}. In \bibinfo{booktitle}{\emph{Proc. ACM SIGMOD.}} \bibinfo{pages}{97--108}.
\newblock
\href{https://doi.org/10.1145/2213836.2213848}{doi:\nolinkurl{10.1145/2213836.2213848}}


\bibitem[Zhang et~al\mbox{.}(2019)]%
        {zhang_table2vec_2019}
\bibfield{author}{\bibinfo{person}{Li Zhang}, \bibinfo{person}{Shuo Zhang}, {and} \bibinfo{person}{Krisztian Balog}.} \bibinfo{year}{2019}\natexlab{}.
\newblock \showarticletitle{{Table2Vec}: {Neural} {Word} and {Entity} {Embeddings} for {Table} {Population} and {Retrieval}}. In \bibinfo{booktitle}{\emph{Proc. 42nd SIGIR. ACM}}. \bibinfo{pages}{1029--1032}.
\newblock
\href{https://doi.org/10.1145/3331184.3331333}{doi:\nolinkurl{10.1145/3331184.3331333}}


\bibitem[Zhang and Balog(2017)]%
        {zhang_entitables_2017}
\bibfield{author}{\bibinfo{person}{Shuo Zhang} {and} \bibinfo{person}{Krisztian Balog}.} \bibinfo{year}{2017}\natexlab{}.
\newblock \showarticletitle{{EntiTables}: {Smart} {Assistance} for {Entity}-{Focused} {Tables}}. In \bibinfo{booktitle}{\emph{Proc. 40th SIGIR. ACM}}. \bibinfo{pages}{255--264}.
\newblock
\href{https://doi.org/10.1145/3077136.3080796}{doi:\nolinkurl{10.1145/3077136.3080796}}


\bibitem[Zhang and Balog(2018)]%
        {zhang_ad_2018}
\bibfield{author}{\bibinfo{person}{Shuo Zhang} {and} \bibinfo{person}{Krisztian Balog}.} \bibinfo{year}{2018}\natexlab{}.
\newblock \showarticletitle{Ad {Hoc} {Table} {Retrieval} using {Semantic} {Similarity}}. In \bibinfo{booktitle}{\emph{Proc. {WWW}}}. \bibinfo{pages}{1553--1562}.
\newblock
\href{https://doi.org/10.1145/3178876.3186067}{doi:\nolinkurl{10.1145/3178876.3186067}}


\bibitem[Zhang and Balog(2021)]%
        {Zhang2021SemanticTR}
\bibfield{author}{\bibinfo{person}{Shuo Zhang} {and} \bibinfo{person}{Krisztian Balog}.} \bibinfo{year}{2021}\natexlab{}.
\newblock \showarticletitle{Semantic Table Retrieval using Keyword and Table Queries}.
\newblock \bibinfo{journal}{\emph{ACM Trans. Web}}  \bibinfo{volume}{15} (\bibinfo{year}{2021}), \bibinfo{pages}{11:1--11:33}.
\newblock


\bibitem[Zhao et~al\mbox{.}(2024)]%
        {Chat2Data}
\bibfield{author}{\bibinfo{person}{Xinyang Zhao}, \bibinfo{person}{Xuanhe Zhou}, {and} \bibinfo{person}{Guoliang Li}.} \bibinfo{year}{2024}\natexlab{}.
\newblock \showarticletitle{Chat2Data: An Interactive Data Analysis System with RAG, Vector Databases and LLMs}.
\newblock \bibinfo{journal}{\emph{Proc. VLDB Endow.}} \bibinfo{volume}{17}, \bibinfo{number}{12} (\bibinfo{year}{2024}), \bibinfo{pages}{4481–4484}.
\newblock
\href{https://doi.org/10.14778/3685800.3685905}{doi:\nolinkurl{10.14778/3685800.3685905}}


\bibitem[Zhu et~al\mbox{.}(2019)]%
        {zhu_josie_2019}
\bibfield{author}{\bibinfo{person}{Erkang Zhu}, \bibinfo{person}{Dong Deng}, \bibinfo{person}{Fatemeh Nargesian}, {and} \bibinfo{person}{Renée~J. Miller}.} \bibinfo{year}{2019}\natexlab{}.
\newblock \showarticletitle{{JOSIE}: {Overlap} {Set} {Similarity} {Search} for {Finding} {Joinable} {Tables} in {Data} {Lakes}}. In \bibinfo{booktitle}{\emph{Proc. ACM SIGMOD.}} \bibinfo{pages}{847--864}.
\newblock
\href{https://doi.org/10.1145/3299869.3300065}{doi:\nolinkurl{10.1145/3299869.3300065}}


\end{thebibliography}

\balance

\end{document}